\documentclass{article}
\usepackage{a4wide,epsfig}
\usepackage{amssymb,cite}

\newtheorem{conjecture}{Conjecture}

\def\dps{\displaystyle}

\def\i{{\rm i}}
\def\e{{\rm e}}
\def\o{{\rm o}}
\def\be{\begin{equation}}
\def\ee{\end{equation}}
\def\bea{\begin{eqnarray}}
\def\eea{\end{eqnarray}}
\begin{document}

\renewcommand{\theequation}{\arabic{section}.\arabic{equation}}

\title{Magic\footnote{\tt The art of producing illusions or tricks that fool or deceive an audience [Webster's Dictionary of American English].}~ in the spectra of the XXZ quantum chain with boundaries at
  $\Delta=0$ and $\Delta=-1/2$.}
\renewcommand{\thefootnote}{\arabic{footnote}}
\author{Jan de Gier$^1$, Alexander Nichols$^2$, Pavel Pyatov$^3$, and Vladimir Rittenberg$^2$\\[5mm] {\small\it
    $^1$Department of Mathematics and Statistics, The University of
  Melbourne, VIC 3010, Australia}\\
{\small\it$^{2}$Physikalisches Institut, Universit\"at Bonn,
  Nussallee 12, Bonn, Germany } \\
{\small\it $^3$Bogoliubov Laboratory of Theoretical Physics, JINR
  141980 Dubna, Moscow Region, Russia \&}\\[-1mm]
{\small\it Max Planck Institute for Mathematics,  Vivatsgasse 7, D-53111,
  Bonn, Germany}}
\date{}
\maketitle

\footnotetext[1]{\tt degier@ms.unimelb.edu.au}
\footnotetext[2]{\tt pyatov@thsun1.jinr.ru}
\footnotetext[3]{\tt nichols@th.physik.uni-bonn.de}
\footnotetext[4]{\tt vladimir@th.physik.uni-bonn.de }

\begin{abstract}
We show that from the spectra of the $U_q (sl(2))$ symmetric XXZ spin-1/2 finite
quantum chain at $\Delta=-1/2$~($q=e^{\pi i/3}$) one can obtain the spectra of certain XXZ quantum chains with diagonal and non-diagonal boundary conditions. Similar
observations are made for $\Delta=0$~($q=e^{\pi i/2}$). In the finite-size
scaling limit the relations
among the various spectra are the result of identities satisfied by known
character functions. For the finite chains the origin of the remarkable spectral
identities can be found in the representation theory of one and two boundaries
Temperley-Lieb algebras at exceptional points. Inspired by these observations
we have discovered other spectral identities between chains with different
boundary conditions.
\end{abstract}

\section{Introduction}
\setcounter{equation}{0}
Finding the spectrum of the spin one-half XXZ quantum chain with non-diagonal
boundaries is still an open problem. We think that this problem is of
special interest because it brings together in a new way the Bethe
Ansatz and the representation theory of some associative algebras.

The XXZ chain with very special diagonal boundary terms has an $U_q(sl(2))$
quantum group symmetry \cite{PasqS90}. This chain is, from an algebraic point of view, simple as it
can be written in terms of the generators of the Temperley-Lieb (TL) algebra. 

The addition of a single boundary term to the $U_q(sl(2))$ chain can be described
using the one-boundary Temperley-Lieb (1BTL)
algebra {\cite{Levy1,Levy2,MartS94,MartS00}. In contrast to the TL algebra the 1BTL now depends on two
parameters. At certain exceptional points (called ``critical values'' in \cite{MartinW00}) this
algebra becomes non-semisimple and possesses indecomposable representations. A third parameter, which
is absent in the algebra, enters as a coefficient in the integrable
Hamiltonian.

As shown in \cite{NichRG05} this general one-boundary 
Hamiltonian has exactly the same spectrum as the XXZ Hamiltonian with purely diagonal
boundary conditions \cite{AlcBBBQ87}. It also has the same spectrum as a
loop Hamiltonian defined on a $2^L$ dimensional space of link patterns
\cite{Gier02}.

All of these Hamiltonians can be written in terms of the 1BTL algebra
using three different representations. These representations are equivalent
except at the exceptional points of the 1BTL algebra. At these
 exceptional points, although the
spectra remain the same, degeneracies appear and the three Hamiltonians have different Jordan cell
structure and therefore describe different physical problems\cite{NichRG05}.

Before proceeding we comment briefly on relations between various algebras we are considering
here. The TL algebra is a quotient of the type $A$ Hecke algebra. The 1BTL algebra contains a TL sub-algebra.
In turn, it is a quotient algebra of type $B$ Hecke algebra, which is a
quotient of affine Hecke algebra (see \cite{Ram1,Ram2} and references therein).

The situation of non-diagonal boundary conditions at both ends of the chain
is more complicated. In addition to the anisotropy parameter, one has five boundary
parameters. As noticed in \cite{deGier:2003iu} the Hamiltonian
can be written in terms of the generators of the two-boundary Temperley-Lieb
algebra \cite{Mitra,Gier02}. This algebra depends on $\Delta$ and three
boundary parameters only. The structure and representation
theory of the two-boundary Temperley-Lieb (2BTL) algebra is essentially
unknown. 

In
\cite{CaoLSW03,Nepomechie:2002xy,Nepomechie:2003vv,Nepomechie:2003ez,deGier:2003iu} the
Bethe Ansatz equations for the case of non-diagonal boundaries was written
only when the boundary parameters satisfy a particular constraint. The surprising fact is that this constraint involves \emph{only} the parameters which
enter the 2BTL algebra and not the coefficients in the Hamiltonian. 

In this paper we
shall make a conjecture for the location of the exceptional points of the 2BTL
algebra and explain how this was obtained. At the exceptional
points the algebra becomes non-semisimple and one finds representations which are
indecomposable. These exceptional points are \emph{exactly} the points at which Bethe Ansatz equations were
written
\cite{CaoLSW03,Nepomechie:2002xy,Nepomechie:2003vv,Nepomechie:2003ez,deGier:2003iu}.
An explanation of this remarkable observation is still missing.

The main aim of this paper is to show some intriguing relations between the spectra
of several Hamiltonians describing the XXZ quantum chain for a finite number
of sites with various boundary terms for $\Delta=-1/2$ and
$\Delta=0$. Some
insight into these relations can be gained from the TL, 1BTL and 2BTL algebras. We hope that the existence of relations of this
kind may bring some new light into the unsolved problem of writing the Bethe Ansatz
equations for arbitrary boundary terms.

We would like to mention that in the case of $\Delta=-1/2$, the Hamiltonians
considered here have the same spectra (but necessarily not the same Jordan cell
structures) as the Hamiltonians describing Raise and Peel stochastic models of
fluctuating interfaces with different sources at the boundaries
\cite{Pyatov04}. In these models the boundary parameters have a simple
physical interpretation.

The paper is organized as follows. In Section \ref{se:TL} we define the TL, 1BTL and 2BTL
algebras and give their representations in terms of XXZ quantum chains. We also
give the relations which define the exceptional points. In sections \ref{se:spectra-pi3} and
\ref{se:spectra-pi2} we give conjectures relating 
different spectra at some fixed values of
the boundary parameters. The magic mentioned in the title of this paper is described
here. These conjectures are based on exact diagonalizations
at a low number of sites given in \ref{se:charpols-pi3} and
\ref{se:charpols-pi2}. In section \ref{se:Bethe} we give proofs and generalizations of several of these
conjectures using the Bethe Ansatz. In sections \ref{se:cft0} and
\ref{se:cft-2} we discuss the finite-size scaling limit. In this limit the
conjectures for finite chains become identities between characters of the
$c=0$ and $c=-2$ conformal field theories. In section
\ref{se:Symmetries} we comment on the appearance of extra symmetries in finite
chains for $\Delta=-1/2$. Conclusions and open questions are in section \ref{se:conc}.
\section{Temperley Lieb algebras and XXZ chains with boundary terms}
\setcounter{equation}{0}
\label{se:TL}

Following \cite{deGier:2003iu,Gier02,NichRG05}, we summarize the relations
between the Temperley-Lieb algebra and its extensions and XXZ quantum chains
with boundaries. As is going to be shown in detail in Section 3, these
relations can help to explain part of the magic observed in the spectra of the
quantum chains seen in Appendices A and B.

We start be defining the algebras. The Temperley-Lieb (TL) algebra is
generated by the unity and the set of $L-1$ elements $e_i$, $i = 1,\ldots,L-1$,
subject to the relations:
\bea \label{eqn:TL}
e_i e_{i\pm 1} e_i&=&e_i \nonumber \\
e_i e_j& =& e_j e_i \quad \quad |i-j|>1\\
e_i^2&=&(q+q^{-1})~e_i\nonumber
\eea
with $q=\exp (\i\gamma)$.
The one-boundary TL algebra (1BTL) (also called blob algebra \cite{MartS94}) is
defined by the generators $e_i$ and a new generator $e_0$. It has the following
additional relations:
\bea \label{eqn:BoundaryTL1}
e_1 e_0 e_1&=&e_1 \nonumber \\
e_0^2&=&\frac{\sin \omega_-}{\sin(\omega_-+\gamma)} e_0 \\
e_0 e_i&=& e_i e_0 \quad \quad i>1 \nonumber
\eea
Notice that the new parameter $\omega_-$ is defined up to a multiple of $\pi$. It was
shown by Martin and Saleur \cite{MartS94} that when for a given value of the bulk
parameter $\gamma$, the boundary parameter $\omega_-$ takes one of the values:
\bea \label{eqn:boundaryTLcritical}
\omega_- = k \gamma + \pi {\mathbb Z}
\eea
with $k$ integer $|k|<L$, the algebra becomes non-semisimple and hence possesses
indecomposable representations. We shall call the values of $\omega_-$ which satisfy
(\ref{eqn:boundaryTLcritical}), exceptional (in
\cite{MartinW00} they are called `critical', a term which
in physics might bring other associations). One should note that 1BTL can have
exceptional points even when $q$ is generic. However when we are in the
exceptional cases (\ref{eqn:boundaryTLcritical}) \emph{and} $q$ is a root of
unity the indecomposable structure is much richer (called `doubly critical' in
\cite{MartinW00}). The focus of this paper will be on such points.

The two-boundary TL algebra (2BTL) is defined by the generators $e_0$, $e_i$
$(i=1,\ldots, L-1)$, and a new generator $e_L$ subject to the
supplementary relations
\bea \label{eqn:BoundaryTL2}
e_{L-1} e_L e_{L-1}&=&e_{L-1} \nonumber \\
e_L^2&=&\frac{\sin \omega_+}{\sin(\omega_++\gamma)} e_L \\
e_{L} e_i&=& e_i e_{L} \quad \quad i < L-1 \nonumber
\eea
and
\be
I_L J_L I_L = b I_L
\label{eq:IJI}
\ee
where
\be
\renewcommand{\arraystretch}{3}
\begin{array}{ll}
\dps I_{2n} = \prod_{x=0}^{n-1} e_{2x+1} & \dps J_{2n} = \prod_{x=0}^{n}
e_{2x} \\
\dps I_{2n+1} = \prod_{x=0}^{n} e_{2x} & \dps J_{2n+1} =
\prod_{x=0}^{n} e_{2x+1}.
\end{array}
\label{eq:idempot}
\ee
Notice that the boundary generators $e_0$ and $e_L$ enter differently in the
expressions of $I_L$ and $J_L$ for $L$ even and odd.
The 2BTL algebra has one bulk parameter $\gamma$ and three boundary parameters
$\omega_{\pm}$ and $b$. It is convenient to use instead of the parameter $b$
another parameter $\theta$ defined by the relations:
\be
b =
\renewcommand{\arraystretch}{2.2}
\left\{ \begin{array}{ll}
\dps \frac{\cos\theta
  -\cos(\gamma+\omega_-+\omega_+)}{2\sin(\gamma+\omega_-)\sin(\gamma+\omega_+)} & {\rm for\; even}\;L \\
\dps \frac{\cos\theta
  +\cos(\omega_--\omega_+)}{2\sin(\gamma+\omega_-)\sin(\gamma+\omega_+)} & {\rm for\; odd}\;L.
\end{array}\right.
\label{eq:defb}
\ee
Based on explicit analysis for small values of $L$, we conjecture that
for a given value of the bulk parameter $\gamma$, the 2BTL algebra has
exceptional points where the 2BTL is non-semisimple if the boundary
parameters $\omega_{\pm}$ and $\theta$ satisfy the relations: 
\be
\left\{ \begin{array}{ll}
\pm\theta = (2k+1)\gamma  + \omega_- + \epsilon~ \omega_+ + 2 \pi {\mathbb Z} & {\rm for}\;{\rm even}\;L,\\
\pm\theta = 2k\gamma + \omega_- + \epsilon~
\omega_+ +\pi + 2 \pi {\mathbb Z}  & {\rm for}\;{\rm odd}\;L,
\end{array}\right.
\label{eq:2bexceptional}
\ee
where $\epsilon =\pm 1$ and $k$ is an integer $|k|<L/2$. Our conjecture is
based on the following facts. 

For small values of $L$ we have found zeros of the determinant of the Gram
matrix also called the discriminant (see \cite{DK}, p.112) 
in two representations of the 2BTL
algebra. Namely, we checked the cases $L=2,3,4$ in the loop representation of
\cite{deGier:2003iu} and $L=2,3$ in the regular representation. The latter
check, 
although more laborious, is a criterium of non-semisimplicity (see for
example exercise 6, p.115 of \cite{DK}). The values given in
(\ref{eq:2bexceptional}) are the exceptional points which depend on $b$. We
also found other exceptional points which were independent of $b$. The non-semisimplicity of the
algebra implies the existence of indecomposable representations.

The algebras TL, 1BTL and 2BTL have a $2^L \times 2^L$ representation which can be
given in terms of Pauli matrices:
\bea \label{eqn:TLgenerators}
e_i= \frac12 \left\{ \sigma^x_i \sigma^x_{i+1} + \sigma^y_i
\sigma^y_{i+1} - \cos \gamma \sigma^z_i \sigma^z_{i+1}  + \cos \gamma
+ \i \sin \gamma \left(\sigma^z_i - \sigma^z_{i+1} \right) \right\}
\eea
for $i = 1,\ldots,L-1$, and
\bea
e_0&=&-\frac12 \frac{1}{\sin(\omega_-+\gamma)}\left( \i \cos \omega_-
\,\sigma_1^z + \sigma_1^x - \sin \omega_- \right), \label{eqn:TLbgenerators_a}\\
e_L&=& -\frac12 \frac{1}{\sin(\omega_++\gamma)}\left( -\i \cos \omega_+
\,\sigma_L^z + \cos\theta \,\sigma_L^x + \sin \theta\, \sigma_L^y - \sin
\omega_+ \right), \label{eqn:TLbgenerators_b}
\eea
As explained in \cite{NichRG05} in the case of the 1BTL, at the exceptional
points, besides the representation given by
(\ref{eqn:TLgenerators})--(\ref{eqn:TLbgenerators_a}) there are other
nonequivalent,  $2^L\times 2^L$ representations of the same algebra. The
same is true for the 2BTL at the exceptional points given in
(\ref{eq:2bexceptional}).

We define three different Hamiltonians which if we use the representation
(\ref{eqn:TLgenerators})--(\ref{eqn:TLbgenerators_b}) have the same
bulk terms with an anisotropy parameter $\Delta = - \cos\gamma$ but
different boundaries:
\bea
H^{\rm T} &=& \sum_{i=1}^{L-1} (1-e_i), \label{eq:hamT}\\
H^{\rm 1T} &=& a_-(1-e_0) + H^{\rm T} ,\label{eq:ham1T}\\
H^{\rm 2T} &=& a_+(1- e_L) + H^{\rm 1T}. \label{eq:ham2T}
\eea
It is convenient to parameterize the coefficients $a_+$ and $a_-$ as follows:
\bea \label{eqn:Definitionofa}
a_\pm= \frac{2 \sin \gamma \sin(\omega_\pm+\gamma)}{\cos \omega_\pm +
  \cos \delta_\pm}.
\eea
We have introduced in the definitions (\ref{eq:hamT})--(\ref{eq:ham2T}) of the Hamiltonians
constant terms such that for $\gamma=\pi/3$, $\omega_\pm=-2\pi/3$ and
$b=1$, the three Hamiltonians describe stochastic processes and
therefore their ground-state energies are equal to zero for any
system size \cite{PearceRGN02,Pyatov04}.

The Hamiltonian $H^{\rm T}$ has special diagonal boundary conditions and is known to
be $U_q(sl(2))$ symmetric \cite{PasqS90}. The Hamiltonian $H^{\rm 1T}$ has the most general
boundary condition at one side of the chain but a fixed diagonal boundary
condition at the other end of the chain:
\bea \label{eqn:Hnd}
H^{1T}&=& \frac{\sin \gamma}{\cos \omega_- +\cos \delta_-}\left( \i \cos
  \omega_- \sigma_1^z + \sigma_1^x  - \sin \omega_- \right) + \frac{2 \sin \gamma \sin(\omega_-+\gamma)}{\cos \omega_- +
  \cos \delta_-} \nonumber\\
&&-\frac12 \left\{ \sum_{i=1}^{L-1} \left( \sigma^x_i \sigma^x_{i+1} +
    \sigma^y_i \sigma^y_{i+1} - \cos \gamma \sigma^z_i \sigma^z_{i+1} -2 + \cos
    \gamma \right) + \i \sin \gamma \left(\sigma^z_1 - \sigma^z_L \right)
\right\}~~~
\eea
In \cite{NichRG05} it was shown that,
remarkably, the Hamiltonian (\ref{eq:ham1T}) has the same spectrum as the Hamiltonian:
\bea \label{eqn:Hd}
H^{\rm d}&=&-\frac12 \left\{ \sum_{i=1}^{L-1} \left( \sigma^x_i
\sigma^x_{i+1} + \sigma^y_i \sigma^y_{i+1} - \cos \gamma \sigma^z_i
\sigma^z_{i+1} -2 + \cos \gamma \right) \right.\nonumber\\
&&\left.+\sin \gamma \left[\tan \left(\frac{\omega_-+\delta_-}{2}\right) \sigma_1^z +
    \tan \left(\frac{\omega_--\delta_-}{2} \right)\sigma_L^z  +2\frac{
      \sin \omega_- - 2\sin(\omega_-+\gamma)}{\cos
      \omega_- +\cos \delta_-} \right] \right\}.~~~~~~
\eea
$H^{\rm d}$ has diagonal boundary terms only and its spectrum was studied since a long
time using the Bethe Ansatz \cite{AlcBBBQ87}. The Hamiltonian $H^{\rm d}$ commutes with
\bea \label{eqn:Sz}
S^z =\frac12 \sum_{i=1}^L \sigma_i^z.
\eea
and therefore for generic values of parameters one would not expect any
degeneracies to occur. However at exceptional points (\ref{eqn:boundaryTLcritical}) the 1BTL algebra gives rise to
degeneracies in $H^{1T}$. As the spectrum of $H^{\rm 1T}$ and $H^{\rm d}$ is
identical this \emph{implies} degeneracies in $H^{\rm d}$ at these points \cite{NichRG05}.

If one examines the
coefficients of $\sigma^z_1$ and $\sigma^z_L$ in (\ref{eqn:Hd}) there is no obvious
difference between the parameters $\omega_-$ and $\delta_-$. We know however that this
is not the case in the Hamiltonian $H^{\rm 1T}$: $\omega_-$ enters the algebra whereas
$\delta_-$ does not. Some observations related to the properties of the spectrum
of $H^{\rm d}$ can be explained \cite{NichRG05} like the absence of $\delta_-$ in the
finite-size scaling limit
\cite{AlcBGR89} and the fact that for finite chains, at the exceptional points, one
observes degeneracies and some energy levels are $\delta_-$ independent and some are
not \cite{AlcBGR89}.

Using (\ref{eqn:TLgenerators})--(\ref{eqn:TLbgenerators_b}),
(\ref{eq:ham2T}), and 
(\ref{eqn:Definitionofa}), $H^{\rm 2T}$ has the following expression:
\bea
H^{\rm 2T} &=& -\frac12 \left\{\sum_{i=1}^{L-1} \left( \sigma^x_i \sigma^x_{i+1} + \sigma^y_i
\sigma^y_{i+1} - \cos \gamma \sigma^z_i \sigma^z_{i+1} -2 + \cos \gamma
\right) \right.
\nonumber\\
&& \hspace{-9.5mm}{} -
\frac{2\sin\gamma}{\cos\omega_++\cos\delta_+}\left( -\frac{\i}{2} (\cos \omega_+ -\cos\delta_+)
\,\sigma_L^z + \cos\theta \,\sigma_L^x + \sin \theta\, \sigma_L^y - \sin
\omega_+ +2\sin(\omega_++\gamma)\right)\nonumber\\
&&\left.\hspace{-10mm}{}- \frac{2\sin\gamma}{\cos\omega_-+\cos
  \delta_-}\left( \frac{\i}{2} (\cos \omega_--\cos\delta_-) 
\,\sigma_1^z + \sigma_1^x - \sin \omega_- +2\sin(\omega_-+\gamma)\right)\right\}.
\label{eq:defham2T}
\eea
One should keep track of the role of the parameters in (\ref{eq:defham2T}). Only the
parameters $\gamma$, $\omega_\pm$ and $\theta$ enter the 2BTL algebra not the $\delta_\pm$
parameters. If we take into account all the parameters, $H^{\rm 2T}$ gives the most
general XXZ model with boundaries.

Not much is known yet about the spectrum of $H^{\rm 2T}$. One reason for
this lack of understanding is the absence of Bethe Ansatz equations
for generic values of the parameters. However, such equations were derived at
a subset of the exceptional points, those with $\epsilon=1$ \cite{CaoLSW03,Nepomechie:2002xy,Nepomechie:2003vv,deGier:2003iu} and
those with $\epsilon=-1$ in the case of $L$ odd
\cite{deGier:2003iu}. It was further noticed that one needs two sets of
Bethe Ansatz equations to describe the complete spectrum
\cite{Nepomechie:2003ez,deGier:2003iu}.

In the next section, we are going to show that for $\gamma = \pi/3$
and $\gamma = \pi/2$, for certain boundary parameters the spectra of the
Hamiltonian $H^{\rm T}$ for even and odd number of sites give all the
energy levels observed in the Hamiltonians $H^{\rm 1T}$ and $H^{\rm
  2T}$. In section \ref{se:Bethe} we shall prove generalizations of some of
these conjectures.
\section{Spectra with magic. The case $q=e^{i \pi/3}$}
\setcounter{equation}{0}
\label{se:spectra-pi3}

\subsection{The open and one-boundary chains.}
\label{se:spectra-pi3-1b}

In this section we are going to take
\be
e_j^2 = e_j,\quad (j = 0,1,\ldots,L).
\ee
This implies $\gamma=\pi/3$ and $\omega_{\pm}=-2 \pi/3$.

We start by considering the spectra of the Hamiltonian $H^{\rm T}$,
defined in (\ref{eq:hamT}) with $\gamma=\pi/3$, which
describes a stochastic process with open boundaries \cite{PearceRGN02}. It
follows that the ground-state energy is zero for any number of sites.

$H^{\rm T}$ is $U_q(sl(2))$ symmetric. Its spectrum can be computed using the
Bethe Ansatz in the spin basis \cite{AlcBBBQ87} or in the link pattern basis
\cite{deGier:2003iu,NichRG05}. The advantage of the latter basis, is that $H^{\rm T}$ has a block
triangular form and that in order to compute the spectrum, one can disregard
the ``off-diagonal'' blocks and be left with the ``diagonal'' ones only. To a
given value of the spin $S$ there are $2S+1$ identical diagonal blocks.

In \ref{se:charpols-pi3} we present the characteristic polynomials
which give the spectra of $H^{\rm T}$ for 
different sizes of the system and different values of $S$. We can notice
that there are supplementary degeneracies which occur because
$q=e^{i\pi/3}$ is a root of unity. These degeneracies are well understood \cite{PasqS90}. The partition function for a system of size $L$ and spin
$S$ and the total partition function are defined as follows:
\bea
Z_{L,S}^{\rm T}(z) &=& \sum_{i} z^{E_i} \\
Z_{L}^{\rm T}(z) &=& \sum_{S} (2S+1) Z_{L,S}^{\rm T}(z)
\label{eq:psum}
\eea
We next consider the Hamiltonian $H^{\rm 1T}$ taking $\omega_-=
-2\gamma = -2\pi/3$ in the
1BTL (\ref{eqn:BoundaryTL1}) and $a_-=1$ (i.e. $\delta_-=\pi$). $H^{\rm 1T}$ describes again
a stochastic process \cite{PearceRGN02}. It is helpful to use
the spectral equivalence between $H^{\rm 1T}$ and $H^{\rm d}$ (see (\ref{eqn:Hd})). Since $S^z$ given by (\ref{eqn:Sz}), commutes
with $H^{\rm d}$, in \ref{se:charpols-pi3} we give for different sizes $L$, the
characteristic
polynomials for different eigenvalues $m$ of $S^z$. Notice that we did
not need to use other polynomials than those used already for
the $U_q(sl(2))$ symmetric chain. We denote by $Z_{L,m}^{\rm 1T}(z)$ the partition function
for a system of size $L$ and charge $m$ and the total partition function by:
\bea 
Z_{L}^{\rm 1T}(z) &=& \sum_{m \in {\mathbb Z}} Z_{L,m}^{\rm 1T}(z)
\eea
\begin{conjecture}
\label{conj:pi3_1}
The following identities hold for finite chains:
\bea
Z^{\rm 1T}_{L,m} &=&
\sum_{n\geq 0}\left\{ Z^{\rm T}_{L+1,(3/2+m+3n)}+ Z^{\rm T}_{L,(m+3n)}\right\},
\;\; \mbox{for~} m\geq 0\ ,\label{eq:pi3_1a}
\\[2mm]
Z^{\rm 1T}_{L,-m} &=&
\sum_{n\geq 0}\left\{ Z^{\rm T}_{L+1,(1/2+m+3n)}+ Z^{\rm T}_{L,(2+m+3n)}\right\},
\;\; \mbox{for~} m\geq -1/2\ .\label{eq:pi3_1b}
\eea
\end{conjecture}
The fact that these identities involve systems of size $L$ and $L+1$ will be
discussed further in \cite{ChicoVladimir}. This conjecture implies for the total partition functions:
\be
Z^{\rm 1T}_L = {1\over 3}(Z^{\rm T}_{L+1}+ Z^{\rm T}_L)\ . \label{eq:pi3_2}
\ee
In (\ref{eq:pi3_1a}) and (\ref{eq:pi3_1b}) $n$ spans all integer
values such that the spin does not exceed the value $L/2$.

These identities were checked using data up to $L=11$ (not included in \ref{se:charpols-pi3}). In the finite-size scaling limit, as
shown in (\ref{eq:chi1T-chiTm}) the identities 
amount to obtain the Gauss model from the $U_q(sl(2))$ symmetric partition
functions \cite{BaueS89}.

How can we understand the identities
(\ref{eq:pi3_1a})--(\ref{eq:pi3_2})? We first notice that
for $\gamma = \pi/3$, $\omega_- = -2\pi/3$  the 1BTL algebra is at an
exceptional point (see (\ref{eqn:boundaryTLcritical})). Then, since $e_0^2=e_0$ the 1BTL algebra has a
quotient $e_0=1$. If we take $e_0=1$ in the one-boundary Hamiltonian (\ref{eq:ham1T}) 
it becomes the Temperley-Lieb Hamiltonian  (\ref{eq:hamT}).
This may explain why one part of the spectrum of
$H^{\rm 1T}$ with $L$ sites comes from the spectrum of 
$H^{\rm T}$ with $L$ sites. On the other hand, if
$e_0$ is subject to the supplementary
condition $e_0e_1e_0 = e_0$, $H^{\rm 1T}$ becomes $H^{\rm T}$ with $L+1$
sites. This may explain why another part of the spectrum of $H^{1T}$ with $L$
sites comes from the spectrum of $H^{\rm T}$ with $L+1$ sites\footnote{The existence of these quotients also explains relations between
properties of stationary states of various raise and peel
models\cite{Pyatov04}.}. This suggests how one might generalize Conjecture
\ref{conj:pi3_1} to the case $a_-\ne 1$ (i.e. $\delta_- \ne \pi$) keeping 
$\omega_-=-2\gamma=-2 \pi/3$ unchanged. Let $Z_L^{\rm 1T}(\delta_-,z)$ be the partition
function of $H^{1T}$ in this case. 

We also consider a new Hamiltonian (describing again a stochastic process \cite{PearceRGN02}):
\be
H^{\rm T} (\delta_-) = a_-(1-\tilde{e}_0) + H^{\rm T},
\label{eq:HTdelta}
\ee
where $\tilde{e}_0$ and $e_i$ ($i=1, \cdots, L-1$) are generators of a TL algebra with $L$ generators. Obviously, $H^{\rm T} (\delta_-)$ is $U_q(sl(2))$
symmetric. Let $Z_L^{\rm T}(\delta_-,z)$ be the total partition function
given by the spectrum of $H^{\rm T} (\delta_-)$.
\begin{conjecture}
\label{conj:pi3_2}
The following identity holds for finite chains:
\be
Z_L^{\rm 1T}(\delta_-,z) = \frac13\left(Z_L^{\rm T}(z) + Z_{L+1}^{\rm T}(\delta_-,z)\right).
\ee
\end{conjecture}
This identity was checked up to $L=6$. There are several
implications of this conjecture. Firstly, if $L$ is even and assuming
that all levels except the ground-state of $H^{\rm T}(\delta_-)$ depend on $\delta_-$, out of the
$2^L$ levels, $p+1$ levels--including the ground-state--will be $\delta_-$
independent and $2p$ will be $\delta_-$ dependent. Here $p= (2^L-1)/3$. If $L$ is
odd, and $p=(2^{L-1} - 1 )/3$, then out of the $2^{L+1}$ levels, $1+2p$ are
$\delta_-$ independent and $4p+1$ are $\delta_-$ dependent. In principle, this
result can also be obtained using the methods developed in \cite{NichRG05}.
The observation that in the spectrum of $H^{\rm d}$ there are levels independent of
$\delta_-$ is known for a long time \cite{AlcBGR89,Bela04}, the fact that this
unusual behaviour of the energy levels is related to exceptional points of
the 1BTL algebra is new (see also \cite{NichRG05}). Moreover, barring accidental
coincidences of energy levels occurring for special values of $a_-$,
the degeneracies seen in the $H^{\rm d}$ chain can be obtained from the known
degeneracies of the $U_q(sl(2))$ symmetric chains $H^{\rm T}$
($\delta_-$) (they are $\delta_-$ independent).

Another interesting consequence of the Conjecture \ref{conj:pi3_2} is
that it gives the finite-size scaling limit of the spectrum of the Hamiltonian
(\ref{eq:HTdelta}) which, to our knowledge is unknown. Since in the
finite-size scaling limit $Z_L^{\rm 1T}(\delta_-,z)$ is $\delta_-$ independent
\cite{AlcBGR89}, from Conjecture \ref{conj:pi3_2} so is $Z_L^{\rm
  T}(\delta_-,z)$.

It is convenient to divide the spectrum of $H^{\rm 1T}$ for $\delta_-=\pi$ and $L$ even into two groups
according to the value of $m$. We denote these two groups by by
$0_{\e}$
and $1/3_{\e}$. The group $0_\e$ contains all the states in the sectors $m = 3k$ and
$m=3k+2$ where $k$ is an integer. The group $1/3_\e$ contains the states with
$m = 3k+1$. For $L$ odd, the two groups denoted by $0_\o$ and $1/3_\o$ contain the
same states with $m$ replaced by $\tilde{m}$ where $\tilde{m} = 1/2 -
m$. In \ref{se:charpols-pi3} for each
value of system size $L$ the two groups are separated. The partition functions
given by the energy levels of the four groups are denoted by
$Z_L^{0_{\e,\o}}(z)$, $Z_L^{1/3_{\e,\o}}(z)$. These partition functions are going to be used when we
consider the two-boundary case and in section \ref{se:cft0}.

\subsection{The two-boundary chain}
\label{se:spectra-pi3-2b}
We turn now to the Hamiltonian $H^{\rm 2T}$ (see (\ref{eq:ham2T}) or (\ref{eq:defham2T})). We
  take $\omega_+=\omega_-= -2\pi/3$ and $a_-=a_+=1$
($\delta_-=\delta_+ = \pi  $). We did not consider other values of $a_-$ and $a_+$

In order to fix the Hamiltonian $H^{\rm 2T}$ we have to specify the values of $b$
in the 2BTL. Firstly, we take $b=1$. This choice makes the Hamiltonian
$H^{\rm 2T}$
describe a stochastic process \cite{Pyatov04} and therefore the ground-state energy
is equal to zero for any number of sites. 

The spectra of the quantum chain for different sizes of the system are
shown in \ref{se:charpols-pi3}. A closer look at the characteristic
polynomials shows that the energy levels which appear for $L$ sites, are contained
in the one-boundary chains of size $L$ and $L+1$ and hence in the open chains of size $L$, $L+1$ and $L+2$. This is not entirely
surprising since we are at exceptional points of the 2BTL algebra
(see (\ref{eq:2bexceptional})). As in the case
of the one-boundary chain one can take quotients in the algebra
$e_0 = e_L=1$ or conversely to promote a boundary generator of the 2BTL
to a generator of the 1BTL algebra. We first separate the states in
$0_\e$ and $0_\o$ into two groups
\bea
Z_L^{0_{\e}} &=& Z_L^{\e,I} + Z_L^{\e,II} \\
Z_L^{0_{\o}} &=& Z_L^{\o,I} + Z_L^{\o,II} 
\eea
where
\bea
Z_L^{\e,I} = \sum_{m\in \mathbb{Z}} Z_{L,3m}^{1T},\qquad
Z_L^{\e,II} = \sum_{m\in \mathbb{Z}} Z_{L,3m+2}^{1T},
\label{eq:ZI-IIa}
\\
Z_L^{\o,I} = \sum_{m\in \mathbb{Z}} Z_{L,3m+1/2}^{1T},\qquad
Z_L^{\o,II} = \sum_{m\in \mathbb{Z}} Z_{L,3m+3/2}^{1T}.
\label{eq:ZI-IIb}
\eea
The data presented in \ref{se:charpols-pi3}, suggest
the following conjecture:
\begin{conjecture}
\label{conj:pi3_5}
The total partition function $Z_L^{b=1}(z)$ 
for the Hamiltonian $H^{\rm 2T}$ for $b=1$ and $L$ sites can be written in terms of
partition functions of the one-boundary chain:
%
\bea
Z_{L=2r}^{b=1}(z) &=& Z_{L=2r}^{\e,I}(z) +
Z_{L=2r+1}^{\o,II}(z) = Z_{L=2r}^{\e,II}(z) +
Z_{L=2r+1}^{\o,I}(z), \label{eq:Z2b=1a}\\
Z_{L=2r-1}^{b=1}(z) &=& Z_{L=2r-1}^{\o,I}(z) +
Z_{L=2r}^{\e,II}(z) = Z_{L=2r-1}^{\o,II}(z) +
Z_{L=2r}^{\e,I}(z). \label{eq:Z2b=1b}
\eea
\end{conjecture}
This conjecture was checked up to $L=8$.

We take now $b=0$. This implies taking $\theta= \pm\pi$ in the quantum
chain (\ref{eq:defham2T}).
The value $b=0$ might look ``natural'', but as
we are going to notice shortly, there is more to this choice.

The spectra for $b=0$ and different lattice sizes are shown in
\ref{se:charpols-pi3}. Keeping the notation introduced at the end of the previous section:
\begin{equation}
\label{eq:Z1/3}
Z_{L=2r}^{1/3_\e} =\sum_{m\in{\mathbb Z}} Z^{\rm 1T}_{L,3m+1}, \qquad
Z_{L=2r-1}^{1/3_\o}=\sum_{m\in {\mathbb Z}} Z^{\rm 1T}_{L,3m-1/2},
\end{equation}
one observes that the following
conjecture is compatible with the data: 
\begin{conjecture}
\label{conj:pi3_6}
The total partition function $Z_L^{b=0}(z)$ 
for the Hamiltonian $H^{\rm 2T}$
for $b=0$ and $L$ sites can be written in terms of
partition functions of the one-boundary chain:
\bea
Z_{L=2r}^{b=0}(z) &=& Z_{L=2r}^{1/3_\e}(z) +
Z_{L=2r+1}^{1/3_\o}(z), \label{eq:Z2b=0a}\\
Z_{L=2r-1}^{b=0}(z) &=& Z_{L=2r-1}^{1/3_\o}(z) +
Z_{L=2r}^{1/3_\e}(z). \label{eq:Z2b=0b}
\eea
\end{conjecture}
This conjecture was checked up to $L=8$. In section \ref{se:Bethe} we shall
prove generalizations of conjectures $3$ and $4$ relating spectra of $H^{1T}$
and $H^{2T}$ for arbitrary values of $\omega_-$ and $\delta_-$ keeping $(\pm
\theta-\omega_-)$ fixed. These are always exceptional points of the
2BTL algebra (\ref{eq:2bexceptional}).  

We have shown that for the one and two-boundary chains with proper
boundary conditions, the energy levels can all be found in the open chain.
Should we look for other values of $b$? Probably not since for $\omega_{\pm}=-2
\gamma$, $\gamma =\pi/3$, one can see from (\ref{eq:2bexceptional}) that there
are no other exceptional points.
\section{Spectra with magic. The case $q=e^{i\pi/2}$.}
\setcounter{equation}{0}
\label{se:spectra-pi2}

\subsection{The open and one-boundary chain.}
\label{se:spectra-pi2-1b}
The case $q=e^{i\pi/2}$ is special since one can find the spectrum and
the wavefunctions of the XX model with the most general boundary conditions
without using the Bethe Ansatz (see \cite{BilsW99} and \cite{Bils00} and references
therein).

We are going to show, that similar to the case $q=e^{i\pi/3}$, magic exists
and can be again related to the TL algebra and its extensions. Throughout
this section we are going to take
\be
e_j^2 = 0,\quad (j = 0,1,\ldots,L).
\ee
This implies $\gamma=\pi/2,~ \omega_- = \omega_+ = -\pi$. (We have fixed the parameters of the
TL and 1BTL algebra and three out of the four parameters of the 2BTL
algebra).

We start with the $U_q(sl(2))$ symmetric Hamiltonian (\ref{eq:hamT}). The
characteristic polynomials for different sizes and spin sectors $S$ are given
in \ref{se:charpols-pi2}.

Next, we consider $H^{\rm 1T}$, defined in (\ref{eq:ham1T}), in which we take $a_-= 1$ ($\delta_-=\pi$).
Using the spectral equivalence of $H^{\rm 1T}$ and $H^{\rm d}$ (see
(\ref{eqn:Hd})) in \ref{se:charpols-pi2} we
give the characteristic polynomials for different values of $m$ (the eigenvalues of
$S_z$). We have to keep in mind that $\omega_-=-\pi$ is an exceptional value of
the 1BTL algebra (see (\ref{eqn:boundaryTLcritical})).

One notices the following two identities:
\be
Z^{\rm 1T}_{L,m}(z) = Z^{\rm 1T}_{L,-m}(z),
\label{eq:pi2_3}
\ee
and
\be
Z^{\rm 1T}_{L,m}(z) = \sum_{n\geq 0}Z^{\rm T}_{L+1,1/2+m+2n}(z)\quad
(m\geq -1/2).
\label{eq:pi2_4}
\ee
from which it follows that:
\be
\sum_{n\geq 0} Z_{L,2n}^{\rm T}(z) = \sum_{n\geq 0} Z_{L,2n+1}^{\rm
  T}(z),\qquad {\rm for}\;L\;{\rm even}
\label{eq:pi2_2}
\ee
Here $Z_{L,S}^{\rm T}$ is the partition function in the spin sector $S$ for a system
size $L$ of $H^{\rm T}$ and $Z_{L,m}^{\rm 1T}$ is the partition function in the sector $m$ for a
system of size $L$ of $H^{\rm 1T}$. The relations
(\ref{eq:pi2_2})--(\ref{eq:pi2_4}) can be proven for any system size \cite{Isaev}.

From (\ref{eq:pi2_4}) one can show that
\be
Z_{L}^{\rm 1T}(z) = \frac12 Z_{L+1}^{\rm T}(z).
\label{eq:pi2_5}
\ee
where $Z_{L}^{\rm 1T}(z)$ is the total partition function of the open chain in which one
takes into account the multiplicity of each spin $S$ sector, and $Z_{L}^{\rm T}(z)$
is the total partition function for the one-boundary chain obtained by
summing over the values of $m$ ( $-L/2\leq m \leq L/2$).

We want to check if, as observed for the case
$q=e^{i \pi/3}$, one cannot extend the relation (\ref{eq:pi2_5}) to the case
when $a_- \ne 1$ (i.e. $\delta_- \neq \pi$) in
(\ref{eq:ham1T}).

We consider the $U_q(sl(2))$ invariant Hamiltonian (\ref{eq:HTdelta}) and compare it's
spectrum with that of $H^{\rm 1T}$, this brings us to a new conjecture
\begin{conjecture}
\label{conj:pi2_1}
The following identity holds for finite chains:
\be
Z_L^{\rm 1T}(\delta_-,z) = \frac12 Z^{\rm T}_{L+1}(\delta_-,z)
\label{eq:pi2_6}
\ee
where $Z^{\rm T}_{L+1}(\delta_-,z)$ is the total partition function for
$H^{\rm T}(\delta_-)$ with $L$ sites and $Z^{\rm 1T}_{L}(\delta_-,z)$ is
the total partition function for $H^{\rm 1T}$ for $L$ sites.
\end{conjecture}
This conjecture was checked up to $L=5$ sites. The existence of this relation should again be related to the fact that
one is at an exceptional point of the 1BTL algebra.

Before proceeding, let us pause for a moment and look again at the relation
(\ref{eq:pi2_5}) in order to illustrate how little the spectra tell us about the
physical problem. $H^{\rm T}$ for $q$ a root of unity has, as is well known, indecomposable Jordan cells,
$H^{\rm d}$ is Hermitian and therefore is fully
diagonalizable. Moreover, if one adds a dummy site to the $L$-site
Hamiltonian $H^{\rm d}$ (a zero fermionic mode)
the spectrum of the Hamiltonian in the larger vector space is precisely
the one of $H^{\rm T}$ with $L+1$ sites. On the other hand, $H^{\rm 1T}$ although it has the
same spectrum as $H^{\rm d}$, has again Jordan cell structures \cite{NichRG05}. These observations
are relevant since, as we are going to see, various characters of $c=-2$
conformal field theory are going to show up in the finite-size scaling
limit (see Section \ref{se:cft-2}) and the observation concerning different theories
having the same partition functions for the finite chains should imply
different conformal field theories related to the same $c=-2$ characters
expressions.

\subsection{The two-boundary chain.}
\label{se:spetra-pi2-2b}
We consider the Hamiltonian $H^{\rm 2T}$ (see (\ref{eq:ham2T}) or
(\ref{eq:defham2T})) in which we take only 
$a_-=a_+=1$ ($\delta_-=\delta_+=\pi$). We employ the same strategy as in the case of $q=e^{i \pi/3}$ and we
are going to get a surprise.

We  first take $b=1$ in the 2BTL algebra (see (\ref{eq:IJI})). This choice is
quite ``natural'' if one has in mind the link pattern representation of the
2BTL, not discussed here (see \cite{deGier:2003iu}). For $L$ even, (the angle $\theta$
in (\ref{eq:defb}) becomes complex) the spectra of $H^{\rm 2T}$ have nothing to do with
those of $H^{\rm T}$ or $H^{\rm 1T}$ (they are not shown in
\ref{se:charpols-pi2}). The situation is 
entirely different if $L$ is odd. In this case, using (\ref{eq:defb}) one finds
$\theta=0$, the Hamiltonian is Hermitian and has a very simple form:
\be
H = \frac12 \left( \sum_{i=1}^{L-1} \sigma_i^x\sigma_{i+1}^x +
\sigma_i^y\sigma_{i+1}^y + \sigma_1^x + \sigma_L^x\right).
\label{eq:pi2_7}
\ee
Before proceeding let us mention a known curious fact
\cite{BilsR00}. For $L$ odd only, the operator 
\be
Y = \frac18 \sum_{i=1}^{L-1} \left[
  \left(1+\left(-\right)^j\right)\sigma_i^x\sigma_{i+1}^y -
  \left(1-\left(-\right)^j\right)\sigma_i^y\sigma_{i+1}^x \right] +
\frac14\left(\sigma_1^y - \sigma_L^y\right),
\ee
commutes with $H$ given by (\ref{eq:pi2_7}). The fact that in the presence of two
boundaries, quantum chains with even and odd number of sites behave so
differently, is a surprise. One possible origin of the surprise can be
found in (\ref{eq:2bexceptional}). If we take $\gamma=\pi/2$, and $\omega_-=\omega_+=-\pi$ in the
equations, for $L$ odd, one finds two solutions: $\theta=0$ and $\theta=\pm\pi$ whereas for $L$
even only one solution: $\theta =\pm\pi/2$. The latter two solutions correspond to
the case $b=0$ which will be discussed soon.

The characteristic polynomials which give the spectrum of $H$ in (\ref{eq:pi2_7})
are shown in \ref{se:charpols-pi2}. Comparing the characteristic polynomials for a
system of size $L=2r-1$ with the characteristic polynomials for the one-boundary case and size $L=2r$, one is led to the following conjecture:
\begin{conjecture}
\label{conj:pi2_2}
The total partition function for the Hamiltonian $H^{\rm 2T}$, $\theta=0$ and $L=2r-1$
sites is related to the charge even sector ($m =2n$) of $H^{\rm 1T}$ with $L=2r$
sites:
\be
Z_{L=2r-1}^{b=1}(z) = \sum_{n \in {\mathbb Z}} Z^{\rm 1T}_{L=2r,2n}(z).
\label{eq:pi2_conj2}
\ee
\end{conjecture}

We consider $b=0$ in the 2BTL algebra. We start with $L$ odd. This
corresponds to $\theta=\pm\pi$. The characteristic polynomials are shown
in \ref{se:charpols-pi2}, their expressions suggest the next
conjecture: 
\begin{conjecture}
\label{conj:pi2_3}
The total partition function for the Hamiltonian $H^{\rm 2T}$, $\theta=\pm\pi$ and $L=2r-1$
sites is related to the charge odd sector ($m=2n+1$) of $H^{\rm 1T}$ with $L=2r$
sites:
\be
Z_{L=2r-1}^{b=0}(z) = \sum_{n \in {\mathbb Z}} Z^{\rm 1T}_{L=2r,2n+1}(z).
\label{eq:pi2_conj3}
\ee
\end{conjecture}

We consider again $b=0$, with $L$ even. This corresponds to $\theta = \pm\pi/2$. The
characteristic polynomials given in \ref{se:charpols-pi2}, suggest the conjecture:
\begin{conjecture}
\label{conj:pi2_4}
The total partition function for the Hamiltonian $H^{\rm 2T}$, $\theta=\pm\pi/2$ and $L=2r$
sites is related to $1/2$ of the partition function of $H^{\rm 1T}$ with $L=2r+1$
sites:
\be
Z_{L=2r}^{b=0}(z) = \frac12\sum_{m \in {\mathbb Z}+1/2} Z^{\rm 1T}_{L=2r+1,m}(z).
\label{eq:pi2_conj4}
\ee
\end{conjecture}
These three conjectures were checked up to $L=8$. It is interesting to mention
that the spectra of the three quantum chains $H^{\rm 2T}$
(\ref{eq:pi2_conj2}), (\ref{eq:pi2_conj3}), and (\ref{eq:pi2_conj4})
exhaust the spectra of $H^{\rm 1T}$. This is another argument for not
expecting more exceptional points of the 2BTL algebra for $\gamma=\pi/2,
\omega_{\pm}=-\pi$. 
As will be discussed in Section \ref{se:cft-2}, all the conjectures made above are
valid in the finite-size scaling limit.

In the next section we shall prove generalizations of conjectures $6$,
$7$ and $8$ relating spectra of $H^{1T}$ and $H^{2T}$ for arbitrary values of
$\omega_-$ and $\delta_-$ keeping $(\pm \theta-\omega_-)$ fixed. These are always
exceptional points of the 2BTL algebra (\ref{eq:2bexceptional}). 

\section{Spectral equivalences from the Bethe Ansatz}
\label{se:Bethe}

The Bethe Ansatz equations for the spectrum of $H^{\rm 2T}$ have been written
down only at exceptional points of the 2BTL
\cite{CaoLSW03,Nepomechie:2002xy,Nepomechie:2003vv,deGier:2003iu}. The cases
in this paper are precisely of this kind, and we will use the Bethe
Ansatz solution to further argue conjectures \ref{conj:pi3_5},
\ref{conj:pi3_6} and \ref{conj:pi2_2}-\ref{conj:pi2_4}. In fact, we
will argue a generalization of these conjectures. 

Let us start by briefly stating the results for $H^{\rm 2T}$ at the exceptional
point (\ref{eq:2bexceptional}) with $\epsilon=1$
\cite{CaoLSW03,Nepomechie:2002xy,Nepomechie:2003vv}. To simplify the
presentation we recall from (\ref{eqn:Definitionofa}) that 
\be
a_\pm = \frac{2\sin\gamma \sin(\gamma+\omega_\pm)}{\cos\omega_\pm + \cos\delta_\pm},
\ee
and we will also use the definition
\be
s_\pm = \frac{\sin\omega_\pm}{\sin(\gamma+\omega_\pm)}.
\ee
The eigenvalues of $H^{\rm 2T}$ split into two groups, $E_1^{\rm 2T}(k)$ and $E_2^{\rm
  2T}(k)$, which can be written as
\bea E_1^{2T}(k) &=& a_-(1-s_-) + a_+(1-s_+) +L-1 - \sum_{j=1}^{\lfloor (L-1)/2\rfloor-k}
  \frac{2\sin^2\gamma}{\cos 2u_i - \cos\gamma},
  \label{eq:E2b-1}\\
E_2^{2T}(k) &=& a_- + a_+ +L-1 - \sum_{j=1}^{\lfloor L/2\rfloor+k}
  \frac{2\sin^2\gamma}{\cos 2v_i - \cos\gamma},
  \label{eq:E2b-2}
\eea
%
for all $k\in{\mathbb Z}$ satisfying (\ref{eq:2bexceptional}). The
complex numbers $u_i$ and $v_i$ are solutions of the equations 
\bea
z(u_i)^{2L} &=&
\frac{K_-(u_i-\omega_-)K_+(u_i-\omega_+)}
     {K_-(-u_i-\omega_-)K_+(-u_i-\omega_+)} \prod_{j=1 \atop
       {j\neq i}}^{\lfloor (L-1)/2\rfloor-k} \frac{S(u_i,u_j)}{S(-u_i,u_j)},
       \label{eq:BA2b-1}\\
z(v_i)^{2L} &=&
\frac{K_-(v_i)K_+(v_i)}
     {K_-(-v_i)K_+(-v_i)} \prod_{j=1 \atop {j\neq i}}^{\lfloor L/2\rfloor+k}
     \frac{S(v_i,v_j)}{S(-v_i,v_j)},
     \label{eq:BA2b-2}
\eea
where we have used the following functions,
\bea
z(u) &=& \frac{\sin(\gamma/2+u_i)}{\sin(\gamma/2-u_i)},\qquad S(u,v) =
\cos 2v - \cos(2\gamma+2u),\\
K_\pm(u) &=& \cos \delta_\pm + \cos(\gamma+\omega_\pm
+2u).
\eea

We are going to compare these solutions to the Bethe Ansatz for
$H^{\rm 1T}$ \cite{deGier:2003iu}. As described in \cite{NichRG05}, the
eigenvalues $E^{\rm 1T}$ of $H^{\rm 1T}$ are the same as those of
$H^{\rm d}$ \cite{AlcBBBQ87} and can therefore be grouped into sectors labelled by the eigenvalues $m$
of $S_z$. We give the Bethe Ansatz equations for $H^{\rm 1T}$
explicitly, to emphasize their similarity with
(\ref{eq:E2b-1})--(\ref{eq:BA2b-2}),  
\be E^{1T}(m) = a_-(1-s_-) +L-1 - \sum_{j=1}^{L/2-m}
  \frac{2\sin^2\gamma}{\cos 2\mu_i - \cos\gamma},
  \label{eq:E1b-1}
\ee
and
\be E^{1T}(m) = a_-+L-1 - \sum_{j=1}^{L/2+m}
  \frac{2\sin^2\gamma}{\cos 2\nu_i - \cos\gamma}.
  \label{eq:E1b-2}
\ee
Here, the complex numbers $\mu_i$ and $\nu_i$ satisfy the equations
\bea
z(\mu_i)^{2L} &=&
\frac{K_-(\mu_i-\omega_-)}
     {K_-(-\mu_i-\omega_-)} \prod_{j=1 \atop
       {j\neq i}}^{L/2-m}
     \frac{S(\mu_i,\mu_j)}{S(-\mu_i,\mu_j)},
       \label{eq:BA1b-1}\\
z(\nu_i)^{2L} &=&
\frac{K_-(\nu_i)}
     {K_-(-\nu_i)} \prod_{j=1 \atop {j\neq i}}^{L/2+m}
     \frac{S(\nu_i,\nu_j)}{S(-\nu_i,\nu_j)}.
     \label{eq:BA1b-2}
\eea
We will now describe some connections between the spectra of $H^{\rm 2T}$ and that of
$H^{\rm 1T}$. In particular, we set $a_+=1,~\omega_+=-2\gamma$
(i.e. $\delta_+=\pi$) and take several values for  $\gamma$ and $b$,
see (\ref{eq:defb}) and (\ref{eq:2bexceptional}).  

\subsection{$\gamma=\pi/3$}

We take $\delta_+=\pi$ and $\omega_+=-2\pi/3$ and find
\be \frac{K_+(u+2\pi/3)}{K_+(-u+2\pi/3)} = 1,\qquad
\frac{K_+(u)}{K_+(-u)} =
\left(\frac{\sin(\gamma/2-u)}{\sin(\gamma/2+u)}\right)^2 =
z(u)^{-2}, \label{eq:Kreduc2}
\ee
This will allow us to identify (\ref{eq:BA2b-1}) and
(\ref{eq:BA2b-2}) with (\ref{eq:BA1b-1}) and (\ref{eq:BA1b-2}) either
with the same system size $L$ or with $L$ replaced by $L+1$. We will
consider this correspondence in detail for $b=1$ and $b=0$. 

\begin{itemize}
\item $b=1$, $L$ even 

From (\ref{eq:defb}) we find that $b=1$ implies $\theta=\pm(\pi+\omega_-)$, and
it follows from (\ref{eq:2bexceptional}) that the spectrum of $H^{\rm
  2T}$ is completely described by (\ref{eq:E2b-1})--(\ref{eq:BA2b-2})
for \emph{each} choice of $k$ with $k=2\bmod 3$. 

Let $Z^{b=1}_L(\omega_-,\delta_-,z)$ be the total partition function
of the Hamiltonian $H^{\rm 2T}$ with $b=1$,  
where the right boundary is fixed as above but with arbitrary values of
$\delta_-$ and $\omega_-$ in the left boundary term.
Denote by $Z_{L=2r}^{\e,I({\rm or}~II)}(\omega_-,\delta_-,z)$,  
$Z_{L=2r+1}^{\o,I({\rm or}~II)}(\omega_-,\delta_-,z)$,
$Z_{L=2r}^{1/3_\e}(\omega_-,\delta_-,z)$ and 
$Z_{L=2r+1}^{1/3_\o}(\omega_-,\delta_-,z)$ the  
partition functions defined as in (\ref{eq:ZI-IIa}),
(\ref{eq:ZI-IIb}), (\ref{eq:Z1/3}) (see also discussion at the end of
section \ref{se:spectra-pi3-1b}) for the Hamiltonian $H^{\rm 1T}$ with
arbitrary values of the left boundary parameters $\omega_-$ and $\delta_-$.

The eigenvalue (\ref{eq:E2b-1}) and the Bethe Ansatz equations 
(\ref{eq:BA2b-1}) are identical to (\ref{eq:E1b-1}) and
(\ref{eq:BA1b-1}) by identifying $m=k+1 \equiv 0\bmod 3$. Similarly,
the eigenvalue (\ref{eq:E2b-2}) and the Bethe Ansatz  
equations (\ref{eq:BA2b-2}) are identical to (\ref{eq:E1b-2}) and
(\ref{eq:BA1b-2}) with $L$ replaced by $L+1$ and $m=k-1/2\equiv
3/2\bmod 3$. This identification holds for all values of $m$, both
positive and negative, although (\ref{eq:E1b-1}) and
(\ref{eq:BA1b-1}) where derived for $m\geq 0$ and (\ref{eq:E1b-2}) and
(\ref{eq:BA1b-2}) for $m\leq -1/2$. Assuming however that the eigenvalue solutions of
(\ref{eq:E1b-2}) and (\ref{eq:BA1b-2}) for $m\leq -1/2$ can also be obtained from
the ``over the equator'' solutions of (\ref{eq:E1b-1}) and
(\ref{eq:BA1b-1}) with $m\leq -1/2$, and vice versa, we obtain from the Bethe Ansatz equations:

\begin{equation}
Z_{L=2r}^{b=1}(\omega_-,\delta_-,z) = Z_{L=2r}^{\e,I}(\omega_-,\delta_-,z) +
Z_{L=2r+1}^{\o,II}(\omega_-,\delta_-,z), \label{eq:Z2b=1a2}
\end{equation}
This is a generalization of the first equality (\ref{eq:Z2b=1a}) in
conjecture \ref{conj:pi3_5}.

\item $b=1$, $L$ odd 

From (\ref{eq:defb}) we find that $b=1$ implies $\theta=\pm(\pi+\omega_-)$, and
it follows from (\ref{eq:2bexceptional}) that $k\equiv 1\bmod 3$. The
eigenvalue (\ref{eq:E2b-1}) and the Bethe Ansatz equations 
(\ref{eq:BA2b-1}) are identical to in (\ref{eq:E1b-1}) and
(\ref{eq:BA1b-1}) by identifying $m=k+1/2 \equiv 3/2\bmod
3$ ($k\geq 1$). Similarly, the eigenvalue (\ref{eq:E2b-2}) and the Bethe Ansatz
equations (\ref{eq:BA2b-2}) are identical to (\ref{eq:E1b-2}) and
(\ref{eq:BA1b-2}) with $L$ replaced by $L+1$ and $m=k-1\equiv 0\bmod
3$ ($k\leq -2$). Using the same assumption as above regarding the
``over the equator'' solutions, we thus find,
\begin{equation}
Z_{L=2r-1}^{b=1}(\omega_-,\delta_-,z) 
= Z_{L=2r-1}^{\o,II}(\omega_-,\delta_-,z) +
Z_{L=2r}^{\e,I}(\omega_-,\delta_-,z). \label{eq:Z2b=1b2}
\end{equation}
The second equality of (\ref{eq:Z2b=1b}) in conjecture \ref{conj:pi3_5} is a corollary.  

\item $b=0$, $L$ even

From (\ref{eq:defb}) we find that $b=0$ implies $\theta=\pm(\omega_--\pi/3)$, and
it follows from (\ref{eq:2bexceptional}) that $k\equiv 0\bmod 3$. The
eigenvalue (\ref{eq:E2b-1}) and the Bethe Ansatz equations 
(\ref{eq:BA2b-1}) are identical to in (\ref{eq:E1b-1}) and
(\ref{eq:BA1b-1}) by identifying $m=k+1 \equiv 1\bmod
3$. Similarly, the eigenvalue (\ref{eq:E2b-2}) and the Bethe Ansatz
equations (\ref{eq:BA2b-2}) are identical to (\ref{eq:E1b-2}) and
(\ref{eq:BA1b-2}) with $L$ replaced by $L+1$ and $m=k-1/2\equiv
5/2\bmod 3$, hence $\widetilde{m}=1/2-m=1-k\equiv 1\bmod
3$. 
Therefore, 
\begin{equation}
Z_{L=2r}^{b=0}(\omega_-,\delta_-,z) = Z_{L=2r}^{1/3_\e}(\omega_-,\delta_-,z) +
Z_{L=2r+1}^{1/3_\o}(\omega_-,\delta_-,z), \label{eq:Z2b=0a2}
\end{equation}
Equation (\ref{eq:Z2b=0a}) in conjecture \ref{conj:pi3_6} follows as a corollary. 

\item $b=0$, $L$ odd

From (\ref{eq:defb}) we find that $b=0$ implies $\theta=\pm(\omega_--\pi/3)$, and
it follows from (\ref{eq:2bexceptional}) that $k\equiv 2\bmod 3$. The
eigenvalue (\ref{eq:E2b-1}) and the Bethe Ansatz equations 
(\ref{eq:BA2b-1}) are identical to (\ref{eq:E1b-1}) and
(\ref{eq:BA1b-1}) by identifying $m=k+1/2$ and hence
$\widetilde{m} =\frac12-m = -k \equiv 1\bmod 3$. Similarly, the
eigenvalue (\ref{eq:E2b-2}) and the Bethe Ansatz 
equations (\ref{eq:BA2b-2}) are identical to (\ref{eq:E1b-2}) and
(\ref{eq:BA1b-2}) with $L$ replaced by $L+1$ and $m=k-1\equiv 1\bmod
3$. 
Hence, 
\begin{equation}
Z_{L=2r-1}^{b=0}(\omega_-,\delta_-,z) = Z_{L=2r-1}^{1/3_\o}(\omega_-,\delta_-,z) +
Z_{L=2r}^{1/3_\e}(\omega_-,\delta_-,z). \label{eq:Z2b=0b2}
\end{equation}
Equation (\ref{eq:Z2b=0b}) in conjecture \ref{conj:pi3_6} follows as a corollary. 
\end{itemize}

\subsection{$\gamma=\pi/2$}

We take $\delta_+=\pi$ and $\omega_+=-2\gamma=-\pi$ and find
\be \frac{K_+(u+\pi)}{K_+(-u+\pi)} =
\frac{K_+(u)}{K_+(-u)} =
\left(\frac{\sin(\gamma/2-u)}{\sin(\gamma/2+u)}\right)^2 =
z(u)^{-2}, \label{eq:Kreduc1}
\ee
As before, this allows us to identify (\ref{eq:BA2b-1}) and (\ref{eq:BA2b-2})
with (\ref{eq:BA1b-1}) and (\ref{eq:BA1b-2}) with $L$ replaced by
$L+1$. We will consider this correspondence in detail for $b=1$ and
$b=0$.

\begin{itemize}
\item $b=1$, $L$ odd

From (\ref{eq:defb}) we find that $b=1$ implies $\theta=\pm(\omega_-+\pi)$,
which means that $k$ in (\ref{eq:2bexceptional}) takes on only odd
values. The eigenvalue (\ref{eq:E2b-1}) and the Bethe Ansatz equations
(\ref{eq:BA2b-1}) are identical to (\ref{eq:E1b-1}) and
(\ref{eq:BA1b-1}) by replacing $L$ with $L+1$ and identifying
$m=k+1$. Similarly, the eigenvalue (\ref{eq:E2b-2}) and
the Bethe Ansatz equations (\ref{eq:BA2b-2}) are identical to
(\ref{eq:E1b-2}) and (\ref{eq:BA1b-2}) with $L$ replaced by $L+1$ but
now $m=k-1$. Assuming again that the eigenvalue solutions of
(\ref{eq:E1b-2}) and (\ref{eq:BA1b-2}) for $m\leq -1/2$ can also be obtained from
the ``over the equator'' solutions of (\ref{eq:E1b-1}) and
(\ref{eq:BA1b-1}) with $m\leq -1/2$, and vice versa, we thus conclude 
\be
Z_{L=2r-1}^{b=1}(\omega_-,\delta_-,z) = \sum_{n \in {\mathbb Z}} 
Z^{\rm 1T}_{L=2r,2n}(\omega_-,\delta_-,z).
\label{eq:pi2_conj2-gen}
\ee
Here $Z_{L=2r-1}^{b=1({\rm or}~0)}(\omega_-,\delta_-,z)$  
(respectively, $Z^{\rm 1T}_{L=2r,m}(\omega_-,\delta_-,z)$) 
denotes the
total (resp., the charge $m$ sector) partition functions
of the Hamiltonian $H^{\rm 2T}$ with $b=1({\rm or}~0)$ (resp., $H^{\rm 1T}$), where
$\delta_+=\pi$ and $\omega_+=-2\gamma=-\pi$ are taken, but
values of the left boundary parameters
$\omega_-$ and $\delta_-$ are kept arbitrary.

Conjecture \ref{conj:pi2_2} follows as a corollary.

\item $b=0$, $L$ odd

This case is complementary to the previous case. Namely $b=0$ implies
from (\ref{eq:defb}) that $\theta=\pm\omega_-$, and now $k$ in
(\ref{eq:2bexceptional}) takes on only even values. Using exactly the
same argument as above, we conclude 
\be
Z_{L=2r-1}^{b=0}(\omega_-,\delta_-,z) = \sum_{n \in {\mathbb Z}} Z^{\rm 1T}_{L=2r,2n+1}(\omega_-,\delta_-,z).
\label{eq:pi2_conj3-gen}
\ee
Conjecture \ref{conj:pi2_3} follows as a corollary.

\item $b=0$, $L$ even

From (\ref{eq:defb}) we find that these values imply $\theta=\pm(\omega_--\pi/2)$,
which means that $k$ in (\ref{eq:2bexceptional}) takes on only even
values. Using again the same reasoning as above, it follows that 
\be
Z_{L=2r}^{b=0}(\omega_-,\delta_-,z) = \frac12\sum_{m \in {\mathbb Z}+1/2} Z^{\rm 1T}_{L=2r+1,m}(\omega_-,\delta_-,z).
\label{eq:pi2_conj4-gen}
\ee
Conjecture \ref{conj:pi2_4} follows as a corollary.
 \end{itemize}

We would like to remark that in \cite{Nepomechie:2002xy} for the case
$\theta=\pi$, $\delta_{+}=\delta_-$, $\omega_+=\omega_-$, and
$\gamma=\frac{\pi}{M+1}$ for any positive integer $M$ (also an exceptional
point of the 2BTL algebra (\ref{eq:2bexceptional})), the spectrum of $H^{2T}$
on $L$ sites (for $L$ odd) is related to the spectra of diagonal chains, and
therefore also of $H^{1T}$. 

Let us finally comment on how the relations 
(\ref{eq:Z2b=1a2})--(\ref{eq:Z2b=0b2})
can be interpreted from an algebraic point of view. In the 2BTL algebra when
we have $e_i^2=e_i$ and $e_L^2=e_L$, 
i.e. $\gamma=\pi/3$ and $ \omega_+=-2 \pi/3$, one can perform two types of
quotient. In the first $e_L=1$ and the $L$-site two-boundary Hamiltonian $H^{\rm 2T}$
(\ref{eq:ham2T}) becomes the $L$-site one-boundary Hamiltonian
$H^{1T}$ (\ref{eq:ham1T}). In the second quotient we take $e_Le_{L-1}e_L=e_L$
and therefore the $L$-site two-boundary Hamiltonian $H^{\rm 2T}$
(\ref{eq:ham2T}) becomes the one-boundary Hamiltonian
$H^{1T}$ (\ref{eq:ham1T}) now on $L+1$ sites. These two quotients, and
relations between the Hamiltonians hold for generic values of 
the boundary parameters $\omega_-$ and $\delta_-$. However we found that the
magical connections between the spectra,
(\ref{eq:Z2b=1a2})--(\ref{eq:pi2_conj4-gen}), only occurred at the
non-semisimple points (\ref{eq:2bexceptional}). A proper understanding of this
fact is still missing.

The conclusion of this section is that the spectrum of the 2BTL Hamiltonian
(\ref{eq:ham2T}) in the exceptional cases can be related to one-boundary Hamiltonian
(\ref{eq:ham1T}). Since the finite-size scaling limit of the spectra of the latter is
known \cite{AlcBBBQ87} this allows us to derive the finite size scaling limit of
the 2BTL case. In the next section we shall do this.    
\section{Finite-size scaling limits of the spectra at
  $q=e^{\pi i/3}$ and $c=0$ CFT}
\setcounter{equation}{0}
\label{se:cft0}

In the next two sections we discuss the finite-size scaling
limit of the spectra discussed in Section \ref{se:spectra-pi3} and
\ref{se:spectra-pi2}. The case of finite size scaling of the Hermitian
two-boundary chain, at the exceptional points (\ref{eq:2bexceptional}), was
considered in \cite{NepomechieAhn}. 

We start with some general observations. As is standard in finite size scaling the partition function (\ref{eq:psum}), used in section \ref{se:spectra-pi3} and
\ref{se:spectra-pi2}, has to be substituted by a different one: 
\be
\widetilde{Z}_L(z) = \sum_i z^{\widetilde{E}_i},
\label{eq:psumt}
\ee
with
\be
\widetilde{E}_i = \frac{(E_i-E_0)L}{\pi v},
\ee
where $E_0$ is the ground-state energy, $L$ is the size of the
system and $v$ is the sound velocity \cite{Hamer},
\be
v= \frac{\pi}{\gamma}\sin\gamma.
\ee

Since all the spectra are contained in the $H^{\rm T}$ Hamiltonian,
so are the ground-states. Therefore, for $q=e^{i\pi/(M+1)}$, in
the finite size scaling limit one can use the result of
\cite{PasqS90} which states that the central charge $c$ of
the Virasoro algebra is:
\be
c= 1 - \frac{6}{M(M+1)}.
\label{eq:ccharge}
\ee
We also expect the finite-size scaling limit of the spectrum of $H^{\rm d}$ in
the sector $m$ (eigenvalue of $S^z$ called ``electric'' charge)  to
be given by a free boson field with Dirichlet-Dirichlet boundary
conditions \cite{Saleur98,Affleck}:
\be
\chi_m = z^{\frac{M}{M+1} (m+\alpha)^2 - \frac{1}{4M(M+1)}}P(z),
\label{eq:chiDD}
\ee
where
\be
P(z) = \prod_{n\geq 1} (1-z^n)^{-1},
\ee
and $\alpha$ depends on $\omega_-$ only and not on $\delta_-$
\cite{AlcBGR89}.

In the case of $H^{\rm 2T}$, the situation is more complicated.
One expects, depending on the boundary parameters, the finite-size scaling
spectrum to be given by the free boson field theory either with
Neumann-Neumann boundary conditions or with 
Neumann-Dirichlet boundary conditions. The connection between the conformal field theory and the boundary parameters of
the XXZ Hamiltonian are known only at the decoupling point
$q=e^{i \pi/2}$ \cite{Bils00}. For $q=e^{i \pi/3}$ such a connection
is not known and one of the results of the magic shown in Section
\ref{se:spectra-pi3} is that in four cases (the four exceptional points of the 2BTL
algebra) we will get it. For what we need, it is sufficient to
mention what we expect in the case of Neumann-Neumann
boundary conditions in the sector $\mu$ \cite{Saleur98,Affleck}
\be \label{eqn:chiNN}
\chi_{\mu} =
z^{\frac{M+1}{M}(\mu+\beta)^2
  -\frac{1}{4M(M+1)}}P(z). 
\ee

Here the sectors are specified by the values $\mu$ (the ``magnetic'' charge)
which are the eigenvalues of an operator related to a $U(1)$
symmetry not seen in the XXZ chain and $\beta$ is an
unknown parameter depending on the boundary parameters.

We consider the case $M=2$. We have, using (\ref{eq:ccharge}), $c=0$. In
the continuum, the spectrum of $H^{\rm T}$ in the sector specified
by spin $S$ gives the partition function (we use
(\ref{eq:psumt})) \cite{BaueS89}:
\be
\chi_S^{\rm T}(z) = \lim_{L \rightarrow \infty} \widetilde{Z}_{L,S}^{\rm T}(z) = \left( z^{S(2S-1)/3} - z^{(S+1)(2S+3)/3}\right)P(z).
\label{eq:chT}
\ee
Using the results of \cite{AlcBGR89} one can fix the value of $\alpha$ in
(\ref{eq:chiDD}) and for the sector $m$ of $H^{\rm d}$ one obtains:
\be
\chi_m^{\rm 1T}(z) = \lim_{L \rightarrow \infty} \widetilde{Z}_{L,m}^{\rm 1T}(z)  = 
z^{2(m-1/4)^2/3-1/24}P(z) = z^{m(2m-1)/3}P(z).
\label{eq:ch1T}
\ee
Using (\ref{eq:chT}) and (\ref{eq:ch1T}) one can check the following identity:
\be
\chi^{\rm 1T}_m = \chi^{\rm 1T}_{1/2-m} = \sum_{n\geq 0} \left(
\chi_{m+3n}^{\rm T} + \chi^{\rm T}_{3/2+m+3n}\right),
\label{eq:chi1T-chiTm}
\ee
which corresponds to the finite size scaling limit of (\ref{eq:pi3_1a}) and (\ref{eq:pi3_1b}) in section \ref{se:spectra-pi3}.

We define:
\bea
\chi^{\rm T}_\e(z) &=& \sum_{S\geq 0}(2S+1)\chi_S^{\rm T},\qquad
\chi^{\rm T}_\o(z) = \sum_{S\geq 1/2}(2S+1)\chi_S^{\rm T}, 
\label{eq:chiTeo} \\
\chi^{\rm 1T}_\e(z) &=& \sum_{m\in\mathbb{Z}}\chi_m^{\rm 1T},\qquad
\chi^{\rm 1T}_\o(z) = \sum_{m\in\mathbb{Z}+1/2}\chi_m^{\rm 1T}. \label{eq:chi1Teo}
\eea
where $e$ and $o$ correspond to the $L \rightarrow \infty$ limits for even and
odd length chains respectively.

Using (\ref{eq:chT}), (\ref{eq:ch1T}) and (\ref{eq:chiTeo}),
(\ref{eq:chi1Teo}) one gets: 
\be
\chi^{\rm 1T}_\e(z) = \chi^{\rm 1T}_\o(z) = \frac13 \left( \chi^{\rm
  T}_\e(z) + \chi^{\rm T}_\o(z)\right),
\label{eq:chi1T-chiT}
\ee
which corresponds to the finite size scaling limit of (\ref{eq:pi3_2}).

In order to understand the spectra of $H^{\rm 2T}$ at the
exceptional points it is important to make contact with characters of $N=2$
superconformal field theory. It is useful to first define the
following functions (with $u = y^{2/3}$):
\bea
\chi^{\e,I}(y,z) &=& \sum_{\mu \in {\mathbb Z}} \chi_{3 \mu}^{\rm 1T}(z) y^{3 \mu}
= u^{1/6} \sum_{\mu \in {\mathbb Z}} u^{2{\mu}-1/6}z^{3(2{\mu}-1/6)^2/2-1/24}P(z),\nonumber\\
\chi^{\e,II}(y,z) &=& \sum_{\mu \in {\mathbb Z}} \chi_{3 {\mu}+2}^{\rm 1T}(z) y^{(3{\mu}+2)}
= u^{1/6} \sum_{\mu \in {\mathbb Z}} u^{2{\mu}+7/6}z^{3(2{\mu}+7/6)^2/2-1/24}P(z),\nonumber\\
\chi^{\o,I}(y,z) &=& \sum_{\mu \in \mathbb Z} \chi_{3{\mu}+1/2}^{\rm 1T}(z) y^{(3{\mu}+1/2)}
= u^{1/6} \sum_{\mu \in {\mathbb Z}} u^{2{\mu}+1/6}z^{3(2{\mu}+1/6)^2/2-1/24}P(z),\nonumber\\[-5mm]
\label{eq:pfuncs}\\
\chi^{\o,II}(y,z) &=& \sum_{\mu \in {\mathbb Z}} \chi_{3{\mu}+3/2}^{\rm 1T}(z) y^{(3{\mu}+3/2)}
= u^{1/6} \sum_{\mu \in {\mathbb Z}} u^{2{\mu}+5/6}z^{3(2{\mu}+5/6)^2/2-1/24}P(z),\nonumber\\
\chi^{1/3,\e}(y,z) &=& \sum_{\mu \in {\mathbb Z}} \chi_{3{\mu}+1}^{\rm 1T}(z) y^{(3{\mu}+1)}
= u^{1/6} \sum_{\mu \in {\mathbb Z}} u^{2{\mu}+1/2}z^{3(2{\mu}+1/2)^2/2-1/24}P(z),\nonumber\\
\chi^{1/3,\o}(y,z) &=& \sum_{\mu \in {\mathbb Z}} \chi_{3{\mu}+5/2}^{\rm 1T}(z) y^{(3{\mu}+5/2)}
= u^{1/6} \sum_{\mu \in {\mathbb Z}} u^{2{\mu}+3/2}z^{3(2{\mu}+3/2)^2/2-1/24}P(z),\nonumber
\eea
It was pointed out by Eguchi and Yang \cite{EguY90} (see also \cite{SalW93})
that the Ramond sector of an $N=2$ superconformal field theory for
any value of the central charge is related to a $c=0$ conformal
field theory. Here we consider the  $N=2$ superconformal for the
case $c=1$ only. 

In the Ramond sector, the $N=2$ superconformal algebra has a
super sub-algebra with four generators; $L_0$ (belonging to the
Virasoro algebra), two supercharges $G_0^i$ $(i=1,2)$ and a $U(1)$
charge $T_0$:
\bea
&&\frac12\left\{G_0^i,G_0^j\right\} = \delta_{ij}\left(L_0
-\frac1{24}\right), \qquad \left[T_0,G_0^i\right] = \i \epsilon_{ij}G^j,
\nonumber\\
&&\left[L_0,G_0^i\right] = \left[L_0,T_0\right] =0.
\label{eq:superc}
\eea
We are going to use this observation later in section \ref{se:Symmetries}. 

The Ramond sector has three representations,
$(1/24,+1/6)$, $(1/24,-1/6)$ and $(3/8,1/2)$ (the first figure
indicates the scaling dimension and the second the charge of the
primary fields) \cite{RavaY88}.

If we define $\tilde{L}_0=L_0-1/24$ then the central charge is shifted from
the value $c=1$ to the value zero and the three representations become:
\be (1/24,\pm 1/6) \rightarrow (0,\pm 1/6),\qquad (3/8,1/2) \rightarrow
(1/3, 1/2)
\ee
With this shift in ground state energy the superalgebra (\ref{eq:superc}) becomes the superalgebra of
quantum mechanics with two supercharges. This superalgebra has
two-dimensional representations if the eigenvalues of $\tilde{L}_0$ are
different of zero. If the eigenvalue of $\tilde{L}_0$ is zero, one has a
one dimensional representation if the supersymmetry is unbroken or
a two-dimensional representation if the supersymmetry is broken
(keep in mind that we have two representations of scaling
dimension $1/24$). Moreover the spectrum of $\tilde{L}_0$ is
non-negative.

The characters in any $N=2$ representation $R$ are defined as:
\be
\chi(u,z)=\mathop{Tr}_R \left( u^{T_0} z^{L_0} \right)
\ee
The characters corresponding to the three representations in the
Ramond sector are given by \cite{RavaY88}:
\bea
\chi_{0,\pm 1/6}(u,z) &=& \sum_{\mu \in\mathbb{Z}} u^{ \mu \pm 1/6} z^{3( \mu
  \pm 1/6)^2/2 -1/24}P(z),
\label{eq:R0}\\
\chi_{1/3,1/2}(u,z) &=& \sum_{ \mu \in\mathbb{Z}} u^{\mu + 1/2} z^{3(\mu+
  1/2)^2/2 -1/24}P(z).
\label{eq:R1/3}
\eea

Notice that the expressions of the characters are similar to those
shown in (\ref{eq:pfuncs}) and therefore we expect to see them in the XXZ
model with non-diagonal boundary conditions.

The $N=2$ superconformal algebra has a sub-superalgebra which is the
$N=1$ superconformal algebra. The last one has, in the Ramond
sector, a sub-superalgebra with only one supercharge: 
\be
G_0^2 = L_0-1/24 = \widetilde{L}_0.
\label{eq:G02}
\ee
Being the square of an Hermitian operator $\tilde{L}_0$ in (\ref{eq:G02})
has a non-negative spectrum.

The $N=1$ superconformal algebra has only one representation with scaling
dimension $1/24$ and one representation with scaling dimension
$3/8$. The two representations $(1/24, \pm 1/6)$ remain irreducible
but coincide in the case of $N=1$ superconformal (the charge which
distinguished them in the case $N=2$ is not present in the case
$N=1$). The representation $(3/8,1/2)$ splits into two identical
representations in the case $N=1$. 

The characters in any $N=1$ representation $R$ are now defined as:
\be
\chi(z)=\mathop{Tr}_R z^{L_0}
\ee
The new characters are:
\bea
\chi^{(N=1)}_{1/24} \rightarrow \chi_0^{(N=1)}(z)&=&\sum_{\mu \in\mathbb{Z}}
z^{3(\mu \pm
  1/6)^2/2 -1/24}P(z) \\
\qquad \chi^{(N=1)}_{3/8} \rightarrow \chi_{1/3}^{(N=1)}(z)&=& \sum_{\mu \in\mathbb{Z}}
z^{3(2 \mu +1/2)^2/2-1/24}P(z) 
\eea
We make now the connection between $N=2$ and $N=1$ superconformal
theories and the XXZ chain with boundaries.

We start with the finite-size spectra of $H^{\rm 1T}$. Comparing
the partition functions (\ref{eq:pfuncs}) and the character expressions
(\ref{eq:R0}) and (\ref{eq:R1/3}) one sees that there is mis-match of the charges
and it is not possible to write the partition functions of $L$
even or $L$ odd separately in terms of $N=2$ characters. One is
able to do it however in terms of $N=1$ superconformal characters:
\be
\sum_{m} \chi_m^{\rm 1T}(z) = \chi^{(1)}_0(z) + \chi^{(1)}_{1/3}(z).
\label{eq:ZN=1}
\ee
for both $m$ even and $m$ odd.

If, however we combine the spectra for $L$ even and odd, the sum
of the partition functions can be expressed in terms of $N=2$
characters: 
\be
\sum_{m\in Z} \chi_m^{\rm 1T}y^m + \sum_{m\in Z+1/2} \chi_m^{\rm 1T}y^m =
\chi_{0,+1/6}(u,z) + \chi_{0,-1/6}(u,z) + \chi_{1/3,1/2}(u,z).
\ee
where $u=y^{3/2}$.

We have to keep in mind that we are in the finite-size scaling
limit. However, since the Hamiltonians describe stochastic processes,
the ground-state energy is zero and the spectrum is positive both for
$H^{\rm 1T}$ and $H^{\rm 2T}$ (if $b=1$) and one
can ask which role the superalgebras (\ref{eq:G02}) and (\ref{eq:superc}) can play
for the finite chains. For $H^{\rm 1T}$ for both $L$ even and odd,
the superalgebra (\ref{eq:G02}) does not give anything new. If we combine
the spectra of $L$ even and $L$ odd however, if magic exists, the
superalgebra (\ref{eq:superc}) might become relevant even for finite chains
and we are going to show in section \ref{se:Symmetries} that this is indeed the case. A closer inspection of
equation (\ref{eq:ZN=1}) suggests that if one excludes from the spectra of
the finite chains for $L$ even and odd separately those states
which contribute to the character with scaling dimensions $1/3$,
one has a chance to be able to use the superalgebra (\ref{eq:superc}). This
is due to the fact that the two representations $(0,\pm 1/6)$ remain
irreducible for $N=1$. In the `cleaned' spectra we will see, in the
finite chain for $L$ even and $L$ odd separately, the ground state
as a singlet and the rest of the spectrum has degeneracies which
are multiples of $2$.

We consider now $H^{\rm 2T}$. Conjectures \ref{conj:pi3_5} and
\ref{conj:pi3_6} of section \ref{se:spectra-pi3} 
give us only the spectra but not any assignment of the ``magnetic''
charge $\tilde{m}$ since no $U(1)$ is known (except at the
decoupling point). We can however ``export'' the $U(1)$ known for
$H^{\rm d}$ and define magnetic charges in this way.

For $b=1$ one finds using (\ref{eq:Z2b=1a}), (\ref{eq:Z2b=1b}),
(\ref{eq:pfuncs}) and (\ref{eq:R0}): 
\bea
\widetilde{Z}_{\rm even}^{b=1}(y,z) &\approx& \widetilde{Z}_{\rm odd}^{b=1}(y,z) \approx
\chi^{\e,I}(y,z) + \chi^{\o,II}(y,z) \approx
\chi^{\e,II}(y,z) + \chi^{\o,I}(y,z) \nonumber \\
&\approx& u^{1/6} \chi_{0,\pm 1/6}(y,z).
\eea
where the sign $\approx$ implies equality modulo a redefinition of
charges.

An amusing observation: in spite of the fact that for $b=1$ one
can expect the superalgebra (\ref{eq:superc}) to play a role for finite
chains (the ground state is indeed a singlet), it does not. There
are singlets for energy non-zero.

For $b=0$ one finds using (\ref{eq:Z2b=0a}), (\ref{eq:Z2b=0b}),
(\ref{eq:pfuncs}) and (\ref{eq:R1/3}): 
\be
\widetilde{Z}_{\rm even}^{b=0}(y,z) = \widetilde{Z}_{\rm odd}^{b=0}(y,z) =
\chi^{1/3,\e}(y,z) + \chi^{1/2,\o}(y,z) = u^{1/6} \chi_{1/3,1/2}(y,z).
\ee
The important conclusion of our discussion is that the four
exceptional points of the 2BTL algebra can be related to
representations of $N=2$ superconformal algebra.
\section{Finite-size scaling limits of the spectra at
  $q=e^{i\pi/2}$ and $c=-2$ CFT} 
\label{se:cft-2} 
\setcounter{equation}{0}

From (\ref{eq:ccharge}) one obtains $c=-2$, a case much
studied in the framework of logarithmic conformal field theory
(see \cite{Gaberd03} and \cite{Flohr03} for reviews). We are going
to touch this topic at the end of the section. Firstly we have to
derive the finite-size scaling limit of the identities discussed
in section \ref{se:spectra-pi2}.

From \cite{BaueS89} and \cite{Bils00} we have:
\be
\chi_S^{\rm T}(z) = \left(z^{S(S-1)/2} - z^{(S+2)(S+1)/2}\right) P(z),
\label{eq:char}
\ee
and:
\be
\chi_m^{\rm 1T}(z) = z^{(m^2-1/4)/2} P(z) = \chi_{-m}^{\rm 1T}(z)
\label{eq:1bchar}
\ee
To derive (\ref{eq:1bchar}) we have used (\ref{eq:chiDD}) and the results of
\cite{AlcBGR89}. Equation
(\ref{eq:1bchar}) corresponds to the finite size scaling limit of (\ref{eq:pi2_3}). From
(\ref{eq:char}) and (\ref{eq:1bchar}) we get:
\be
\chi_m^{\rm 1T}(z) = \sum_{n\geq 0} \chi^{\rm T}_{1/2+m+2n},
\label{eq:char1bchar}
\ee
which corresponds to the finite size scaling limit of (\ref{eq:pi2_4}). We define the total
characters for $H^{\rm T}$
\be
\chi_{\e}^{\rm T}(z) = \sum_{S=0}(2S+1)\chi_S^{\rm T},\qquad
\chi_{\o}^{\rm T}(z) = \sum_{S=1/2}(2S+1)\chi_S^{\rm T},
\label{eq:chartot}
\ee
and $H^{\rm 1T}$
\be
\chi_{\e}^{\rm 1T}(z) = \sum_{m\in \mathbb{Z}}\chi_m^{\rm 1T},\qquad
\chi_{\o}^{\rm 1T}(z) = \sum_{m\in \mathbb{Z}+1/2}\chi_m^{\rm 1T},
\label{eq:1bchartot}
\ee
Using (\ref{eq:char1bchar})--(\ref{eq:1bchartot}) we derive the relations:
\be
\chi_\e^{\rm 1T}(z) = \frac12 \chi_\o^{\rm T}(z),\qquad \chi_\o^{\rm 1T}(z) = \frac12 \chi_\e^{\rm T}(z)
\ee
These relations correspond to (\ref{eq:pi2_5}).

We now check if (\ref{eq:pi2_conj2})--(\ref{eq:pi2_conj4}), conjectured for the finite
chains, are valid in the finite size scaling limit. We use here the results
of \cite{Bils00} which give the values of $\beta$ in (\ref{eqn:chiNN}) as a function
of the angle $\theta$ appearing in the expression of $H^{\rm 2T}$. We also show that
for each value of $\theta$ the partition functions in the finite-size
scaling limit are given by one character only of the $c=-2$ theory.

One has:
\bea
\lim_{k\rightarrow\infty} \widetilde{Z}^{b=1}_{L=2r-1} (z) &=&
\sum_{n\in\mathbb{Z}} \chi^{\rm 1T}_{2n} (z) \qquad (\theta=0)\nonumber \\
&=& \sum_{n\in \mathbb{Z}} z^{2n^2-1/8}P(z) = z^{-1/8} \Theta_{0,2}(z) P(z) \\
\lim_{k\rightarrow\infty} \widetilde{Z}^{b=0}_{L=2r-1} (z) &=&
\sum_{n\in\mathbb{Z}} \chi^{\rm 1T}_{2n+1} (z) \qquad (\theta=\pi)\nonumber \\
&=& \sum_{n\in \mathbb{Z}} z^{2(n+1/2)^2-1/8}P(z) = z^{-1/8} \Theta_{2,2}(z) P(z) .\\
\lim_{k\rightarrow\infty} \widetilde{Z}^{b=0}_{L=2r} (z) &=&
\frac12 \sum_{n\in\mathbb{Z}} \chi^{\rm 1T}_{n+1/2} (z) \qquad (\theta=\frac{\pi}{2})\nonumber \\
&=& \sum_{n\in \mathbb{Z}} z^{2(n+1/4)^2-1/8}P(z) = z^{-1/8} \Theta_{1,2}(z) P(z).
\eea
where
\be
\Theta_{\lambda,\kappa}(z) = \sum_{n\in \mathbb{Z}} z^{(2\kappa n+\lambda)^2/4\kappa}.
\ee
In the continuum there is a well-studied example of a $c=-2$ logarithmic
CFT. This theory has an extended $W$-symmetry with only a finite number of
irreducible and indecomposable representations \cite{Gaberd03,Flohr03}. There
are four irreducible fields conventionally written as
$V_0,V_1,V_{-1/8},V_{3/8}$ where the subscript is the conformal
dimension. There are also two reducible but indecomposable modules $R_0$
and $R_1$. The characters ($\chi=Tr z^{L_0}$) are given by:
\bea
\chi_{V_0}&=&\frac{1}{2} z^{-1/8} \left(\Theta_{1,2}(z) + \partial \Theta_{1,2}(z)\right)P(z) \\
\chi_{V_1}&=&\frac{1}{2} z^{-1/8} \left(\Theta_{1,2}(z) - \partial \Theta_{1,2}(z)\right)P(z) \\
\chi_{V_{-1/8}}&=& z^{-1/8} \Theta_{0,2}(z) P(z) \\
\chi_{V_{3/8}}&=&z^{-1/8} \Theta_{2,2}(z) P(z)  \\
\chi_{R}&=& 2 z^{-1/8} \Theta_{1,2}(z) P(z)
\eea
where $\chi_R \equiv \chi_{R_0}=\chi_{R_1}=2 \left( \chi_{V_0} + \chi_{V_1}\right)$ and:
\be
\partial \Theta_{\lambda,\kappa} = \sum_{n\in \mathbb{Z}} (2 \kappa n +
\lambda) z^{(2\kappa n+\lambda)^2/4\kappa}.
\ee
The finite size scaling of the lattice partition functions can now be
identified with the continuum results:
\bea
\lim_{k\rightarrow\infty} \widetilde{Z}^{b=1}_{L=2r-1} (z) &=& \chi_{V_{-1/8}}\\
\lim_{k\rightarrow\infty} \widetilde{Z}^{b=0}_{L=2r-1} (z) &=& \chi_{V_{3/8}}\\
\lim_{k\rightarrow\infty} \widetilde{Z}^{b=0}_{L=2r} (z) &=& \chi_{V_0} +
\chi_{V_1}=\frac{1}{2} \chi_R
\eea
A very relevant observation is that to each of the three exceptional
points of the 2BTL algebra corresponds to a representation of the $c=-2$
theory in the finite size scaling limit.

The finite-size scaling limit of the spectra of $H^{\rm 1T}$ taken separately
for $L$ even and odd can also be expressed in terms of $c=-2$ characters:
\bea
\chi_\e^{\rm 1T}(z) &=& \chi_{-1/8}(z) + \chi_{3/8}(z),\\
\chi_\o^{\rm 1T}(z) &=& \chi_{{\rm R}}(z).
\eea
We know that $H^{1T}$ contains indecomposable representations and one can
explicitly compute wavefunctions using the methods of \cite{Bils00}. Further
work is required to understand properly the continuum limit of this well
defined lattice model.
\section{Symmetries in the $H^{1T}$ and $H^d$ quantum chains for $q=e^{i \pi/3}$}
\label{se:Symmetries}

In this section we are going to discuss the degeneracies occurring in the
spectra of the Hamiltonian $H^{1T}$ given by \ref{eq:ham1T}. First we are going to
take $a_-=1$ ($\delta_- = \pi$). Before we present our conjectures, it is
instructive to consider the cases $L=5$ and $L=6$ and compare the characteristic
polynomials appearing for different values of $m$.

In \ref{se:charpols-pi3}, for each value of $L$, the characteristic polynomials
for different values of $m$ are separated into two groups. For $L$ even, the group
called $0_e$ in Section \ref{se:spectra-pi3-2b}, is given first and contains the characteristic
polynomials for $m=3k$ and $m=3k+2$ ($k \in {\mathbb Z}$); the second group called
$1/3_e$ contains the characteristic polynomials corresponding to $m=3k+1$. For
$L$ odd, we have the group $0_o$ with $m=3k+1/2$ and $m=3k+3/2$ and the group
$1/3_o$ with $m=3k+5/2$.

For $L=5$ the group $0_o$ contains the levels with $m=-5/2,-3/2,1/2$ and $3/2$.
We notice that the ground state (energy zero) appears as a singlet, all
the other states appear as doublets except the level of energy $5$ which appears
as a quadruplet. The group $1/3_o$ ($m=-1/2$ and $5/2$) contains the energy
level $4$ twice, the others as singlets. Notice that the energy level $4$ appears
in both groups.

What happens with the degeneracies if we change the value of $a_-$ ($\delta_-
\ne \pi$) in $H^{1T}$? The degeneracies within each group stay unchanged but the
energy level $4$ which was common to the two groups splits into two values, one for
the doublet in $0_o$ and a second one for the doublet in $1/3_o$. This observation
is true for all values of $L$: the degeneracies within each of the two
groups are independent on $a_-$ and therefore have to be related to
properties of the 1BTL algebra. This is indeed the case. Degeneracies
appearing between both groups are accidental.

The 1BTL algebra has a center containing several elements. These elements,
which can be written as a linear combination of words of the 1BTL algebra,
have the property that they commute with \emph{all} elements in the
algebra. If one specifies the XXZ representation of the 1BTL algebra (\ref{eqn:TLgenerators}) and (\ref{eqn:TLbgenerators_a}) , one obtains operators (centralizers) which commute with the
Hamiltonian $H^{1T}$ (\ref{eq:ham1T}) (\ref{eqn:Hnd}) written in the same representation. At the
exceptional points of the 1BTL algebra (\ref{eqn:boundaryTLcritical}) where the algebra is not semi-simple, the
centralizers which commute with $H^{1T}$ are of Jordan form. It is easy
to check that if an $N \times N$ matrix is a Jordan cell, any
matrix commuting with it (in particular $H^{1T}$) has to have an $N$-fold degeneracy. One
centralizer found by Doikou \cite{Doikou} was discussed in detail in \cite{NichRG05}, the
construction of the centralizers and their properties will be published
elsewhere. The main message is that the use of the centralizers allows to
get all the degeneracies observed in the $H^{1T}$ quantum chain.

Part of the symmetries observed in the groups $0_e$ and $0_o$ can be
understood in the following way. We consider again the $L=5$ example shown
in \ref{se:charpols-pi3}. It is convenient to order the characteristic
polynomials according to the value of $\tilde{m}$. We notice that apart the ground-state
which appears at $\tilde{m}=0$, one sees doublets at $\Delta \tilde{m} = \pm
1$. For other values of $L$ one can define a new ``charge'' such that doublets
occur in the same way. This implies that one obtains representations of the
superalgebra (\ref{eq:G02}). The fact that the spectrum in the sectors $0_e$
and $0_o$ is composed of doublets and
one singlet is contained in the following conjecture:
\begin{conjecture}
\label{conj:symmetries_1}
The partition functions of $H^{1T}$ satisfy the following identity for all
values of $L$ and $\delta$:
\begin{itemize}
\item{$L$ even}
\bea \label{eqn:conjsymmetries_1Leven}
\sum_{n \in {\mathbb Z}} Z^{1T}_{L,3n} (\delta_-) = 1 + \sum_{n \in {\mathbb Z}} Z^{1T}_{L,3n+2}(\delta_-)
\eea
\item{$L$ odd}
\bea
\sum_{n \in {\mathbb Z}} Z^{1T}_{L,3n+1/2} (\delta_-) = 1 + \sum_{n \in {\mathbb
    Z}} Z^{1T}_{L,3n+3/2}(\delta_-)
\eea
\end{itemize}
\end{conjecture}
The total number of states contained in the spaces $Z_{L,3n}$ and $Z_{L,3n+2}$
for $L$ even and $Z_{L,3n+1/2}$ and $Z_{L,3n+3/2}$
for $L$ odd are given by:
\begin{center}
\begin{tabular}{l|cccccccc}
Length of chain: & 1 & 2 & 3 & 4 & 5 & 6 & 7 & 8 \\
\hline
Number of states: & 1 & 3 & 5 & 11 & 21 & 43 & 85 & 171 
\end{tabular}
\end{center}
These are well known as the Jacobsthal numbers which have many interesting
combinatorical properties \cite{Heubach}. This particular way of realising
them has been previously discussed in \cite{Barry}. 

Notice that the sectors $1/3_e$ and $1/3_0$ contain many singlets and
therefore have nothing to do with the superalgebra (\ref{eq:superc}). $H^{1T}$ is compatible
with the superalgebra (\ref{eq:G02}) for any value of $\delta$ in a trivial
way since it describes a stochastic process and therefore the spectrum is
non-negative. We have not tried to find the odd generators of the
superalgebras (\ref{eq:superc}) and (\ref{eq:G02}).

We are going to compare now the spectra for system sizes $L=2p-1$ and $L=2p$
taking again $a_-=1$ and using the tables of \ref{se:charpols-pi3}. We start with the
example $p=3$. We notice that not only the sectors $0_e$ and $0_o$ have common
energy levels but so do the sectors $1/3_e$ and $1/3_o$. Combining the
spectra of the $L=5$ and $L=6$ one obtains only doublets (including the
ground-state!). Some doublets may have the same energy. These observations are
valid for any value of $p$ and for $a_-=1$ only. The super-algebra (\ref{eq:superc}) gives
doublets in the ground-state only if the supersymmetry is spontaneously
broken. The fact that for $a_-=1$, combining the spectra of the sectors $1/3_e$ for $2p$
sites and $1/3_o$ for $2p-1$ sites, one obtains doublets is the basis of the
following conjecture:
\begin{conjecture}
\label{conj:symmetries_2}
The partition functions of $H^{1T}$ for $a_-=1$ only, satisfy the
following identity:
\bea \label{eqn:conjsymmetries_2}
\sum_{n \ge 0} \left( Z^{1T}_{2p,3n+1} + Z^{1T}_{2p-1,3n+5/2}\right) = \sum_{n
  < 0} \left( Z^{1T}_{2p,3n+1} + Z^{1T}_{2p-1,3n+5/2} \right)
\eea
\end{conjecture}
Let us observe that relations between spectra of different sizes
which are only valid for $a_-=1$ are accidental if one thinks of symmetries
coming from the 1BTL algebra. In a different approach \cite{FendSN03} relations
to 
supersymmetry were found for the Hamiltonian $H^d$ (see (\ref{eqn:Hd})) for the
value $a_-=1$. Let us elaborate shortly about the difference between the
relations (\ref{eqn:conjsymmetries_1Leven})-(\ref{eqn:conjsymmetries_2})  presented above and the remarkable construction in \cite{FendSN03}.
Whereas the relations (\ref{eqn:conjsymmetries_1Leven})-(\ref{eqn:conjsymmetries_2}) refer to sectors of the XXZ chain
defined by $L$ and $m$, the content of \cite{FendSN03} can be found in the
identities of conjecture \ref{conj:pi3_1}. For example one can notice that the
energy levels in 
$Z^T_{L,m+3n}$ are present in both $Z^{1T}_{L-1,m-3/2}$ and
$Z^{1T}_{L,m}$. Those are the doublets of the supersymmetry of \cite{FendSN03}. The XXZ chain is not
supersymmetric. One can however form another system, different from the above,
of size given again by the Jacobsthal numbers which is supersymmetric and
whose energy levels, but not their degeneracies, coincide with those of the
$U_q(sl(2))$ chain \cite{ChicoVladimir}.

We conclude this section with an amusing observation. If one looks at the
spectra of $H^{2T}$ shown in \ref{se:charpols-pi3} although in the finite-size
scaling limit they give the same characters as those of $H^{1T}$, the
degeneracies are very different. New degeneracies appear and some of the
energy levels coincide between systems of different sizes. This suggests, that
as in the case of the $H^{1T}$ Hamiltonian for which the symmetries were given
by the center of the 1BTL algebra, the center of the 2BTL algebra will play an
important role for understanding the new degeneracies. The way to make a
connection between the spectra of systems of different sizes has also to be understood.  
\section{Conclusion}
\label{se:conc}

We have observed that for the finite open XXZ spin chain with boundaries
at $\Delta=0$ and $\Delta=-1/2$ magic occurs. This magic was discovered
accidentally when several computer outputs were lying on one desk.
 
We have first noticed that the same energy levels appearing in the 
$U_q(sl(2))$ symmetric chain with an even and odd number of sites can be 
seen in an non-$U_q(sl(2))$ invariant XXZ chain with certain diagonal 
boundary conditions. The degeneracies are not the same and energy levels seen 
in chains with an even and odd number of sites get mixed up. The main 
point is that the rules how to obtain the spectra of the diagonal chain
(including the degeneracies) from the spectra of the $U_q(sl(2))$ 
invariant chain are simple (see conjecture \ref{conj:pi3_1} and eqs \ref{eq:pi2_3}-\ref{eq:pi2_5}). The same 
phenomenon appears if we look at the XXZ chain with some non-diagonal boundary
conditions (see conjectures \ref{conj:pi3_5}, \ref{conj:pi3_6}, \ref{conj:pi2_2}, \ref{conj:pi2_3}, and \ref{conj:pi2_4}). 

We can, at least partially, understand the origin of this magic
in the following way. The $U_q(sl(2))$ symmetric XXZ chain is
written in terms of a representation of the Temperley-Lieb algebra. The 
exceptional points of this algebra where the algebra is not semisimple (it 
contains indecomposable representations) occur if $q$ is a root of unity.
This is the case for $q=e^{i \pi/3}$ and $q=e^{i \pi/2}$. The XXZ chain with
diagonal boundary conditions can be expressed in terms of generators of
the one-boundary Temperley-Lieb algebra. This algebra has also exceptional 
points \ref{eqn:boundaryTLcritical} and it is at these exceptional points that the magic was observed.
In the case of the chain with non-diagonal boundary conditions, by accident, 
we have chosen the parameters such that we have exhausted the exceptional points 
of the two-boundary Temperley-Lieb algebra compatible with our choice of 
the boundary terms of the diagonal chain. There are four exceptional 
points in the case $q=e^{i \pi/3}$ and three for $q=e^{i \pi/3}$. For 
the exceptional points one expects special properties of the  
representations of the algebras. It is the interplay of the representations 
which produces the magic. This insight in the problem was possible because we 
were able to find an expression for the exceptional points of the two-boundary 
Temperley-Lieb algebra (see eq.\ref{eq:2bexceptional}).

Once we understood the connection between the quantum chains and the
Temperley-Lieb algebra and its extensions, we were able to extend the
``domain of magic'' beyond our original observations (see conjectures \ref{conj:pi3_2}
and \ref{conj:pi2_1}). Next, we wanted to go beyond numerics, and use 
the Bethe Ansatz in order to check the conjectures. It turned out that 
precisely at the exceptional points of the two-boundary Temperley-Lieb 
algebra (again magic!) one can write the Bethe Ansatz equations. This is 
possible because at the exceptional points, the Hamiltonian has a block 
triangular form, and that in order to compute the spectrum, one can 
disregard the off-diagonal blocks and keep only the diagonal ones. At least in
the cases discussed here, the Bethe Ansatz equations for each block 
coincided with the equations obtained for a block of the Hamiltonian with 
diagonal boundary conditions. In the latter case, the diagonal 
blocks are labeled by the eigenvalues $m$ of $S^z$ (eq. \ref{eqn:Sz}). In this way, 
part of the conjectures were confirmed by the Bethe Ansatz equations (see 
section \ref{se:Bethe}). A look at the Bethe Ansatz equations for more general boundary 
conditions suggested a further extension of the 'magic' 
(see eqs.(\ref{eq:Z2b=1a2})--(\ref{eq:pi2_conj4-gen})). The 
origin of this new extension could also be located using the one and 
two-boundary Temperley-Lieb algebra 
(see comment in the end of sec.5.2). For the points we investigated, under the
assumption that the one-boundary Bethe ansatz is complete, our results imply
that the two-boundary Bethe ansatz is also complete. 
 
The ``magic'' was observed on finite chains. We have extended it to the
finite-size scaling limit. Part of the ``magic'' was confirmed by using
known characters of conformal field theory for the $U_q(sl(2))$ symmetric 
case and for the Gauss model (see sections \ref{se:cft0} and \ref{se:cft-2}).

In our opinion, two remarkable observations have to be mentioned. In the
finite-size scaling limit to each of the exceptional points of the
two-boundary Temperley-Lieb algebra corresponds a representation of a chiral
conformal field theory: $c=0$ in the Ramond sector of a $N=2$ superconformal
for $q=e^{i \pi/3}$. In the case $c=-2$ observed for $q=e^{i \pi/2}$ the 
situation is more subtle since the quantum field theory is less under 
control. One obtains some combinations of a particular $c=-2$ conformal field theory. The origin of 
this mismatch is not clear but can in principle be fixed since for $q=e^{i \pi/2}$ one is at the decoupling point and one can easily derive the continuum theory 
from its lattice realization.  The second observation has to do with Jordan
cell structure. In non-minimal models, character functions don't 
characterize alone the conformal field theory. This phenomenon can be seen 
explicitly in our study. For example, the chains with diagonal boundary 
conditions we considered are Hermitian. They have however the same spectra 
as some chains with non-diagonal boundary conditions on one side of the 
chain, corresponding to another representation of the one-boundary 
Temperley-Lieb algebra \cite{NichRG05}. In this different representation on has Jordan 
cells and therefore a different model.

Finally, let us observe, that understanding the connection between the 
quantum chains and the Temperley-Lieb algebra and its extensions has 
another bonus. The quantum chains $H^{1T}$ and $H^{2T}$ (see
eqs. (\ref{eqn:Hnd}) and (\ref{eq:defham2T}) have in general spectra with no degeneracies. This is not the case at the 
exceptional points of the one and two-boundary Temperley Lieb algebras.
An example of this kind is discussed in detail in section \ref{se:Symmetries} where part of 
the degeneracies are encoded in Conjectures \ref{conj:symmetries_1} and \ref{conj:symmetries_2}.
\section*{Acknowledgment}
JdG and VR are grateful for financial support from the Deutsche
Forschungsgemeinschaft and the Australian Research Council. AN and VR
gratefully acknowledge support from the EU network {\it Integrable
  models and applications: from strings to condensed matter}
HPRN-CT-2002-00325. PP was partially supported by the
Heisenberg-Landau grant INTAS-OPEN-03-51-3350 and by the RFBR grant No. 05-01-01086.

\newpage
\appendix \renewcommand\thesection{Appendix \Alph{section}}
\renewcommand{\theequation}{\Alph{section}.\arabic{equation}}
\setcounter{equation}{0}
\section{Characteristic polynomials in the case $\bf \gamma=\pi/3$}
\label{se:charpols-pi3}
Here we present characteristic polynomials, factorised over the integers, for the Hamiltonians of the open
(i.e. $U_q(sl(2))$ invariant) chains (see (2.9), (2.12)), 
the 1-boundary chains (see (2.13), (2.16)) with $\omega_- =
-2\pi/3$, $a_-=1$ ($\delta_-=\pi$), 
and the 2-boundary chains (see (2.14), (2.19)) for $\omega_{\pm}= -2\pi/3$,
$a_{\pm}=1$ ($\delta_{\pm}=\pi$) in two cases $b=1$ and $b=0$.
In all cases we fix anisotropy  $\Delta=-1/2$ ($q=e^{\pi/3}$).
We collect data for  the chains of sizes $1\leq L\leq 8$
and organize them according to values of spin $S$ for the open chains
(each sector $S$ comes with multiplicity $2S+1$)
and according to values of the charge $m$ for the 1-boundary chains.

We give characteristic polynomials factorized over the integers. For each factor we use its total degree
adding subscripts to distinguish between different factors of the same
total degree. For example $4_a$ and $4_b$ are two different factors of degree 4.
Groups of factors which appear always together are taken in parentheses.
Explicit expressions for factors are presented at the end of the appendix.
\begin{center}
\noindent
\hspace{2mm}
\begin{tabular}{|rr|l|}
\multicolumn{3}{c}{$\bf L=1$} \\
\hline
$S$ & & \multicolumn{1}{|c|}{\rule{0pt}{4mm}open chain} \\
\hline
1/2 & & \rule{0pt}{4mm}$({\bf 1_0})$\\
& & \\
\hline
$m$ & $\tilde{m}$ & \multicolumn{1}{|c|}{ \rule{0pt}{4mm}1-boundary chain} \\
\hline
$1/2$ & $0$ & \rule{0pt}{4mm}$({\bf 1_0})$\\ & & \\ & & \\
$-1/2$ & $1$ &  $({\bf 1_1})$  \\
\hline
\multicolumn{3}{|c|}{ \rule{0pt}{4mm}2-boundary chain, $b=1$ case} \\
\hline
\multicolumn{3}{|c|}{\rule{0pt}{4mm} $({\bf 1_2})$ $({\bf 1_0})$} \\
\hline
\multicolumn{3}{|c|}{\rule{0pt}{4mm} 2-boundary chain, $b=0$ case} \\
\hline
\multicolumn{3}{|c|}{\rule{0pt}{4mm} $({\bf 1_1})^2$} \\
\hline
\end{tabular}
\begin{tabular}{|r|l|}
\multicolumn{2}{c}{$\bf L=2$} \\
\hline
$S$ & \multicolumn{1}{|c|}{\rule{0pt}{4mm}open chain} \\
\hline
0 & \rule{0pt}{4mm}$({\bf 1_0})$\\
1 & $({\bf 1_1})$ \\
\hline
$m$ & \multicolumn{1}{|c|}{ \rule{0pt}{4mm}1-boundary chain} \\
\hline
$0$ & \rule{0pt}{4mm}$({\bf 1_2})$ $({\bf 1_0})$ \\
$-1$ &  $({\bf 1_2})$\\ & \\
$1$ & $({\bf 1_1})$  \\
\hline
\multicolumn{2}{|c|}{ \rule{0pt}{4mm}2-boundary chain, $b=1$ case} \\
\hline
\multicolumn{2}{|c|}{\rule{0pt}{4mm}
$({\bf 1_3})$ $({\bf 1_2})^2$ $({\bf 1_0})$ } \\
\hline
\multicolumn{2}{|c|}{\rule{0pt}{4mm} 2-boundary chain, $b=0$ case} \\
\hline
\multicolumn{2}{|c|}{\rule{0pt}{4mm}
$({\bf 2_a})$ $({\bf 1_2})$$({\bf 1_1})$ } \\
\hline
\end{tabular}

\vskip5mm
\noindent
\begin{tabular}{|rr|l|}
\multicolumn{3}{c}{$\bf L=3$} \\
\hline
$S$ & & \multicolumn{1}{|c|}{\rule{0pt}{4mm}open chain} \\
\hline
1/2 & & \rule{0pt}{4mm}$({\bf 1_2})$ $({\bf 1_0})$\\
3/2 & & $({\bf 1_2})$ \\
& & \\
\hline
$m$ & $\tilde{m}$ & \multicolumn{1}{|c|}{ \rule{0pt}{4mm}1-boundary chain} \\
\hline
$-3/2$ & $2$ & \rule{0pt}{4mm}$({\bf 1_3})$\\
$1/2$ & $0$ & $({\bf 1_3})$ $({\bf 1_2})$ $({\bf 1_0})$ \\
$3/2$ & $-1$ & $({\bf 1_2})$\\ & & \\
$-1/2$ & $1$ &  $({\bf 2_a \cdot 1_2})$  \\
 & & \\
\hline
\multicolumn{3}{|c|}{ \rule{0pt}{4mm}2-boundary chain, $b=1$ case} \\
\hline
\multicolumn{3}{|c|}{\rule{0pt}{4mm}
$({\bf 2_b\cdot 2_c})$ $({\bf 1_3})^2$ $({\bf 1_2})$ $({\bf 1_0})$ }\\
\hline
\multicolumn{3}{|c|}{\rule{0pt}{4mm} 2-boundary chain, $b=0$ case} \\
\hline
\multicolumn{3}{|c|}{\rule{0pt}{4mm}
$({\bf 2_a})^2$ $({\bf 1_4})^2$$({\bf 1_2})^2$ } \\
\hline
\end{tabular}
\begin{tabular}{|r|l|}
\multicolumn{2}{c}{$\bf L=4$} \\
\hline
$S$ & \multicolumn{1}{|c|}{\rule{0pt}{4mm}open chain} \\
\hline
0 & \rule{0pt}{4mm}$({\bf 1_3})$ $({\bf 1_0})$\\
1 & $({\bf 2_a\cdot 1_2})$ \\
2 & $({\bf 1_3})$\\
\hline
$m$ & \multicolumn{1}{|c|}{ \rule{0pt}{4mm}1-boundary chain} \\
\hline
$2$ & \rule{0pt}{4mm}$({\bf 1_3})$  \\
$0$ & $({\bf 2_b\cdot 2_c})$ $({\bf 1_3})$ $({\bf 1_0})$ \\
$-1$ & $({\bf 2_b\cdot 2_c})$   \\ & \\
$1$ &  $({\bf 2_a\cdot 1_2})$$({\bf 1_4})$\\
$-2$ &   $({\bf 1_4})$\\
\hline
\multicolumn{2}{|c|}{ \rule{0pt}{4mm}2-boundary chain, $b=1$ case} \\
\hline
\multicolumn{2}{|c|}{\rule{0pt}{4mm}
$({\bf 2_b\cdot 2_c})^2$ $({\bf 2_d\cdot 1_3\cdot 1_4})$
$({\bf 1_3})$ $({\bf 1_5})^2$ $({\bf 1_0})$} \\
\hline
\multicolumn{2}{|c|}{\rule{0pt}{4mm} 2-boundary chain, $b=0$ case} \\
\hline
\multicolumn{2}{|c|}{\rule{0pt}{4mm}
$({\bf 6\cdot 3_a})$ $({\bf 2_a\cdot 1_2})$
$({\bf 1_4})^4$ } \\
\hline
\end{tabular}

\vspace{10pt}
\noindent
\begin{tabular}{|rr|l|}
\multicolumn{3}{c}{$\bf L=5$} \\
\hline
$S$ & & \multicolumn{1}{|c|}{ \rule{0pt}{4mm} open chain} \\
\hline
\rule{0pt}{4mm}
1/2 & & $({\bf 2_b\cdot 2_c})$ $({\bf 1_0})$ \\
3/2 & & $({\bf 2_b\cdot 2_c})$ \\
5/2 & & $({\bf 1_4})$\\
 & & \\
\hline
$m$ & $\tilde{m}$ & \multicolumn{1}{|c|}{\rule{0pt}{4mm} 1-boundary chain} \\
\hline
\rule{0pt}{4mm}$-5/2$ & $3$&  $({\bf 1_5})$  \\
$-3/2$ & $2$ & $({\bf 2_d\cdot 1_3\cdot 1_4})$ $({\bf 1_5})$ \\
$1/2$ & $0$ & $({\bf 2_b\cdot 2_c})$ $({\bf 2_d\cdot 1_3\cdot 1_4})$$({\bf 1_5})$ $({\bf 1_0})$\\
$3/2$ & $-1$ & $({\bf 2_b\cdot 2_c})$ $({\bf 1_5})$ \\
& & \\ & & \\ & & \\
$-1/2$ & $1$ &  $({\bf 6\cdot 3_a})$ $({\bf 1_4})$ \\
$5/2$ & $-2$&  $({\bf 1_4})$  \\
\hline
\multicolumn{3}{|c|}{ \rule{0pt}{4mm} 2-boundary chain, $b=1$ case} \\
\hline
\multicolumn{3}{|c|}{\rule{0pt}{4mm}
$({\bf 7\cdot 5\cdot 1_5})$ $({\bf 2_b\cdot 2_c})$
$({\bf 2_d\cdot 1_3\cdot 1_4})^2$} \\
\multicolumn{3}{|c|}{
$({\bf 1_5})^4$ $({\bf 1_6})^2$ $({\bf 1_0})$ } \\
\hline
\multicolumn{3}{|c|}{\rule{0pt}{4mm} 2-boundary chain, $b=0$ case} \\
\hline
\multicolumn{3}{|c|}{\rule{0pt}{4mm}
$({\bf 6\cdot 3_a})^2$ $({\bf 3_b\cdot 3_c})^2$
$({\bf 1_4})^2$} \\
\hline
\end{tabular}
\begin{tabular}{|r|l|}
\multicolumn{2}{c}{$\bf L=6$} \\
\hline
$S$ & \multicolumn{1}{|c|}{\rule{0pt}{4mm}open chain} \\
\hline
0 & \rule{0pt}{4mm}$({\bf 2_d\cdot 1_3\cdot 1_4})$ $({\bf 1_0})$ \\
1 & $({\bf 6\cdot 3_a})$ \\
2 & $({\bf 2_d\cdot 1_3\cdot 1_4})$ $({\bf 1_5})$\\
3 & $({\bf 1_5})$\\
\hline
$m$ & \multicolumn{1}{|c|}{ \rule{0pt}{4mm}1-boundary chain} \\
\hline
$3$ & \rule{0pt}{4mm}$({\bf 1_5})$\\
$2$ & $({\bf 2_d\cdot 1_3\cdot 1_4})$ $({\bf 1_6})$ $({\bf 1_5})$ \\
$0$ & $({\bf 7\cdot 5\cdot 1_5})$ $({\bf 2_d\cdot 1_3\cdot 1_4})$
$({\bf 1_6})$\\ &  $({\bf 1_5})$ $({\bf 1_0})$ \\
$-1$ & $({\bf 7\cdot 5\cdot 1_5})$  $({\bf 1_6})$ $({\bf 1_5})$ \\
$-3$ & $({\bf 1_6})$\\
& \\
$1$ & $({\bf 6\cdot 3_a})$ $({\bf 3_b\cdot 3_c})$ \\
$-2$ &  $({\bf 3_b\cdot 3_c})$ \\
\hline
\multicolumn{2}{|c|}{ \rule{0pt}{4mm}2-boundary chain, $b=1$ case} \\
\hline
\multicolumn{2}{|c|}{\rule{0pt}{4mm}
$({\bf 9\cdot 4_a})$
$({\bf 7\cdot 5\cdot 1_5})^2$
$({\bf 4_b\cdot 2_e\cdot 1_6})^2$} \\
\multicolumn{2}{|c|}{$({\bf 2_d\cdot 1_3\cdot 1_4})$
$({\bf 1_5})^2$ $({\bf 1_6})^4$$({\bf 1_0})$} \\
\hline
\multicolumn{2}{|c|}{\rule{0pt}{4mm} 2-boundary chain, $b=0$ case} \\
\hline
\multicolumn{2}{|c|}{\rule{0pt}{4mm}
$({\bf 16\cdot 10\cdot 1_4})$
$({\bf 6\cdot 3_a})$ $({\bf 3_b\cdot 3_c})^4$
$({\bf 1_7})^4$ } \\
\hline
\end{tabular}

\vspace{10pt}
\noindent
\begin{tabular}{|rr|l|}
\multicolumn{3}{c}{$\bf L=7$} \\
\hline
$S~$ & & \multicolumn{1}{|c|}{ \rule{0pt}{4mm} open chain} \\
\hline
\rule{0pt}{4mm}
1/2 & & $({\bf 7\cdot 5\cdot 1_5})$ $({\bf 1_0})$ \\
3/2 & & $({\bf 7\cdot 5\cdot 1_5})$ $({\bf 1_6})$ \\
5/2 & & $({\bf 3_b\cdot 3_c})$\\
7/2 & & $({\bf 1_6})$\\
&&\\
\hline
$m$ & $\tilde{m}$ & \multicolumn{1}{|c|}{\rule{0pt}{4mm} 1-boundary chain} \\
\hline
\rule{0pt}{4mm}
$-5/2$ & $3$ & $({\bf 4_b\cdot 2_e\cdot 1_6})$  \\
$-3/2$ & $2$ & $({\bf 9\cdot 4_a})$ $({\bf 4_b\cdot 2_e\cdot 1_6})$ $({\bf 1_6})$ \\
$1/2$  & $0$ & $({\bf 9\cdot 4_a})$ $({\bf 7\cdot 5\cdot 1_5})$ $({\bf
  4_b\cdot 2_e\cdot 1_6})$\\ 
 & & $({\bf 1_6})$ $({\bf 1_0})$ \\
$3/2$ & $-1$ & $({\bf 7\cdot 5\cdot 1_5})$ $({\bf 4_b\cdot 2_e\cdot 1_6})$ $({\bf 1_6})$ \\
$7/2$ & $-3$ & $({\bf 1_6})$  \\
& & \\
& & \\
& & \\
$-7/2$ & $4$ & $({\bf 1_7})$  \\
$-1/2$ & $1$ & $({\bf 16\cdot 10\cdot 1_4})$ $({\bf 3_b\cdot 3_c})$
$({\bf 1_7})^2$ \\
$5/2$& $-2$ & $({\bf 3_b\cdot 3_c})$ $({\bf 1_7})$  \\
\hline
\multicolumn{3}{|c|}{ \rule{0pt}{4mm} 2-boundary chain, $b=1$ case} \\
\hline
\multicolumn{3}{|c|}{\rule{0pt}{4mm}
$({\bf 20_a\cdot 20_b\cdot 1_8})$
$({\bf 9\cdot 4_a})^2$
$({\bf 7\cdot 5\cdot 1_5})$
 }\\
\multicolumn{3}{|c|}{$({\bf 4_b\cdot 2_e\cdot 1_6})^4$
$({\bf 3_d\cdot 3_e\cdot 1_6})^2$ $({\bf 1_6})^2$ $({\bf 1_8})^3$
$({\bf 1_0})$ } \\
\hline
\multicolumn{3}{|c|}{\rule{0pt}{4mm} 2-boundary chain, $b=0$ case} \\
\hline
\multicolumn{3}{|c|}{\rule{0pt}{4mm}
$({\bf 16\cdot 10\cdot 1_4})^2$ $({\bf 15\cdot 12})^2$
$({\bf 3_b\cdot 3_c})^2$
$({\bf 1_7})^8$} \\
\hline
\end{tabular}
\begin{tabular}{|r|l|}           
\multicolumn{2}{c}{$\bf L=8$} \\                                                              
\hline                                                                                       
$S$ & \multicolumn{1}{|c|}{\rule{0pt}{4mm}open chain} \\                                      
\hline                                                                                        
0 & \rule{0pt}{4mm}$({\bf 9\cdot 4_a})$ $({\bf 1_0})$ \\                           
1 & $({\bf 16\cdot 10\cdot 1_4})$ $({\bf 1_7})$\\                                            
2 & $({\bf 9\cdot 4_a})$ $({\bf 4_b\cdot 2_e\cdot 1_4})$ \\
3 & $({\bf 4_b\cdot 2_e \cdot 1_6})$\\                   
4 & $({\bf 1_7})$\\  
\hline                                                                                        
$m$ & \multicolumn{1}{|c|}{ \rule{0pt}{4mm}1-boundary chain} \\                               
\hline            
$3$ & \rule{0pt}{4mm}$({\bf 4_b\cdot 2_e\cdot 1_6})$ $({\bf 1_8})$\\ 
$2$ & $({\bf 9\cdot 4_a})$ $({\bf 4_b\cdot 2_e\cdot 1_6})$ $({\bf
  3_d\cdot 3_e\cdot 1_6})$ $({\bf 1_8})$\\ 
$0$ & $({\bf 20_a\cdot 20_b\cdot 1_8})$ $({\bf 9\cdot 4_a})$ $({\bf
  4_b\cdot 2_e\cdot 1_6})$ \\
& $({\bf 3_d\cdot 3_e\cdot 1_6})$ $({\bf
  1_8})$ $({\bf 1_0})$ \\
$-1$ & $({\bf 20_a\cdot 20_b\cdot 1_8})$ $({\bf 4_b\cdot 2_e\cdot
  1_6})$ \\
& $({\bf 3_d\cdot 3_e\cdot 1_6})$ $({\bf 1_8})$ \\
$-3$ & $({\bf 3_d\cdot 3_e\cdot 1_6})$ $({\bf 1_8})$\\
$-4$ & $({\bf 1_8})$\\
& \\
$4$ & $({\bf 1_7})$\\
$1$ & $({\bf 16\cdot 10\cdot 1_4})$ $({\bf 15\cdot 12})$ $({\bf
  1_7})^2$ \\
$-2$ & $({\bf 15\cdot 12})$ $({\bf 1_7})$\\  
\hline                                                                                        
\multicolumn{2}{|c|}{ \rule{0pt}{4mm}2-boundary chain, $b=1$ case} \\                         
\hline                                                                                        
\multicolumn{2}{|c|}{\rule{0pt}{4mm}--------
}\\ 
\multicolumn{2}{|c|}{}\\
\hline                                                                                        
\multicolumn{2}{|c|}{\rule{0pt}{4mm} 2-boundary chain, $b=0$ case} \\                         
\hline
\multicolumn{2}{|c|}{\rule{0pt}{4mm} --------} \\
\hline
\end{tabular}                                                                                 
\end{center}

Here
\be
\begin{array}{l}
\begin{array}{lll}
{\bf 1_n} := x-n, &  & \\[2mm]
{\bf 2_a} := x^2 - 4x + 2,\qquad\quad & {\bf 2_b} :=x^2 - 5x + 5,\qquad\quad &
{\bf 2_c} :=x^2 - 7x + 11, \\
{\bf 2_d} :=x^2 - 8x + 13,& {\bf 2_e} :=x^2 - 12x + 34, &
\end{array}
\\[10mm]
\begin{array}{ll}
{\bf 3_a} :=x^3 - 10x^2 + 30x - 26, &
{\bf 3_b} :=x^3 - 14x^2 + 63x - 91,\\
{\bf 3_c} :=x^3 - 16x^2 + 83x - 139, &
{\bf 3_d} :=x^3 - 21x^2 + 144x - 321,\\
{\bf 3_e} :=x^3 - 21x^2 + 144x - 323, \qquad\quad&
\end{array}
\\[10mm]
\begin{array}{l}
{\bf 4_a} :=x^4 - 18x^3 +117x^2 - 324x + 321, \\
{\bf 4_b}:=x^4 - 24x^3 + 212x^2 - 816x + 1154, \\[2mm]
{\bf 5} :=x^5 - 21x^4 + 170x^3 - 661x^2 + 1229x - 867, \\[2mm]
{\bf 6} :=x^6 - 20x^5 + 157x^4 - 610x^3 + 1204x^2 - 1078x + 278, \\[2mm]
{\bf 7} :=x^7 - 28x^6 + 323x^5 - 1983x^4 + 6962x^3 - 13868x^2 + 14323x -
5770, \\[2mm]
{\bf 9} :=x^9 - 45x^8 + 882x^7 - 9872x^6 + 69450x^5 - 317988x^4 +
945869x^3 - 1758591x^2 + \\ 
\hspace{12mm}1849158x - 834632, \\[2mm]
{\bf 10} :=x^{10} - 46x^9 + 933x^8 - 10972x^7 + 82698x^6 - 416480x^5 +
1415230x^4 - \\
\hspace{12mm}3192220x^3 + 4551945x^2 - 3680424x + 1268488, \\[2mm]
{\bf 12} := x^{12} - 75x^{11} + 2562x^{10} - 52701x^9 + 726928x^8
-7081826x^7 +  49954540x^6 - \\
\hspace{12mm}257012522x^5 + 956940353x^4 -
2513874287x^3 + 4421248479x^2 - \\
\hspace{12mm}4672270934x + 2242636033,\\[2mm] 
{\bf 15} := x^{15} - 93x^{14} + 4008x^{13} - 106163x^{12} +
1932458x^{11} - 25600562x^{10} + \\
\hspace{12mm} 254927932x^9 - 1942567842x^8 +
11417435665x^7 - 51744755105x^6 + \\
\hspace{12mm}179295171389x^5 - 466283136174x^4 +
880666793685x^3 - 1139877184096x^2 + \\
\hspace{12mm}903639748800x - 330565630976,\\[2mm]
{\bf 16} :=x^{16} - 76x^{15} + 2670x^{14} -  57512x^{13} + 849351x^{12} -
9109048x^{11} + \\
\hspace{12mm} 73289680x^{10} - 450525464x^9 + 2134046231x^8 -
7794633798x^7 + \\
\hspace{12mm} 21803583759x^6 -  45993980288x^5 + 71220198638x^4
- 77391639144x^3 + \\
\hspace{12mm} 54589655356x^2 -  21469924224x + 3193100216,\\[2mm]
{\bf 20_a} := x^{20} - 107x^{19} + 5390x^{18} - 169909x^{17} +
3757766x^{16} - 61956744x^{15} + \\
\hspace{12mm}789839374x^{14} - 7968451646x^{13} +
64579301106x^{12} - 424317702016x^{11} + \\
\hspace{12mm}2271049739581x^{10} -
9910587278544x^9 + 35165896339844x^8 - \\
\hspace{12mm}100788840091272x^7 +
230719077491798x^6 - 414591247028377x^5 + \\
\hspace{12mm}569799308661865x^4 -
575489965955241x^3 + 400153245868113x^2 - \\
\hspace{12mm}169764861535134x +
32741611046721,\\[2mm]
{\bf 20_b} := x^{20} - 109x^{19} + 5600x^{18} - 180287x^{17} +
4078540x^{16} - 68907116x^{15} + \\
\hspace{12mm}901976466x^{14} - 9365114226x^{13} +
78317050846x^{12} - 532575594652x^{11} + \\
\hspace{12mm}2960318111641x^{10} -
13469643127548x^9 + 50063934003660x^8 - \\
\hspace{12mm}151115303430784x^7 +
366651617570206x^6 - 703734083636583x^5 + \\
\hspace{12mm}1042848523879615x^4 -
1149127999221811x^3 + 885073633040283x^2 - \\
\hspace{12mm}424451387835254x +
95207779114473.
\end{array}
\end{array}
\ee
\newpage
\section{Characteristic polynomials in the case $\bf \gamma=\pi/2$}
\label{se:charpols-pi2}
Here we present characteristic polynomials for the Hamiltonians of the open
(i.e. $U_q(sl(2))$ invariant) chains (see (2.9), (2.12)), 
the 1-boundary chains (see (2.13), (2.16)) with $\omega_-=-\pi$, $a_-=1$ ($\delta_-=\pi$), 
and the 2-boundary chains (see (2.14), (2.19)) for $\omega_{\pm}=-\pi $,
$a_{\pm}=1$ ($\delta_{\pm}=\pi$) and in cases $b=1$ and $b=0$ (for $L$
even we only consider $b=0$). In all cases we fix anisotropy  $\Delta=0$ ($q=e^{\pi/2}$).

As for the previous appendix we collect data for the chains of sizes $1\leq L\leq 8$
and organize them according to values of spin $S$ for the open chains
(each sector $S$ comes with multiplicity $2S+1$)
and according to values of the charge $m$ for the 1-boundary chains. Explicit expressions for factors are presented at the end.

\begin{center}
\noindent
\hspace{2mm}
\begin{tabular}{|r|l|}
\multicolumn{2}{c}{$\bf L=1$} \\
\hline
$S$ & \multicolumn{1}{|c|}{\rule{0pt}{4mm}open chain} \\
\hline
1/2 & \rule{0pt}{4mm}$({\bf 1_0})$ \\ & \\
\hline
$m$ & \multicolumn{1}{|c|}{ \rule{0pt}{4mm}1-boundary chain} \\
\hline
 $1/2$ & \rule{0pt}{4mm}$({\bf 1_1})$\\
$-1/2$ &  $({\bf 1_1})$  \\ & \\
\hline
\multicolumn{2}{|c|}{ \rule{0pt}{4mm}2-boundary chain, $b=1$ case} \\
\hline
\multicolumn{2}{|c|}{\rule{0pt}{4mm} $({\bf 1_1})$ $({\bf 1_3})$ } \\
\hline
\multicolumn{2}{|c|}{\rule{0pt}{4mm} 2-boundary chain, $b=0$ case} \\
\hline
\multicolumn{2}{|c|}{\rule{0pt}{4mm} $({\bf 1_2})^2$} \\
\hline
\end{tabular}
\begin{tabular}{|r|l|}
\multicolumn{2}{c}{$\bf L=2$} \\
\hline
$S$ & \multicolumn{1}{|c|}{\rule{0pt}{4mm}open chain} \\
\hline
0 & \rule{0pt}{4mm}$({\bf 1_1})$ \\
1 & $({\bf 1_1})$ \\
\hline
$m$ & \multicolumn{1}{|c|}{ \rule{0pt}{4mm}1-boundary chain} \\
\hline
 $0$ & \rule{0pt}{4mm}$({\bf 1_1\cdot 1_3})$\\
$1$ &  $({\bf 1_2})$  \\
$-1$ & $({\bf 1_2})$   \\
\hline
\multicolumn{2}{|c|}{ \rule{0pt}{4mm}2-boundary chain, $b=1$ case} \\
\hline
\multicolumn{2}{|c|}{\rule{0pt}{4mm} --------} \\
\hline
\multicolumn{2}{|c|}{\rule{0pt}{4mm} 2-boundary chain, $b=0$ case} \\
\hline
\multicolumn{2}{|c|}{\rule{0pt}{4mm}
$({\bf 2_a})$ $({\bf 1_3})^2$ } \\
\hline
\end{tabular}

\vskip5mm
\begin{tabular}{|r|l|}
\multicolumn{2}{c}{$\bf L=3$} \\
\hline
$S$ & \multicolumn{1}{|c|}{\rule{0pt}{4mm}open chain} \\
\hline
1/2 & \rule{0pt}{4mm}$({\bf 1_1\cdot 1_3})$ \\
3/2 & $({\bf 1_2})$ \\ & \\
\hline
$m$ & \multicolumn{1}{|c|}{ \rule{0pt}{4mm}1-boundary chain} \\
\hline
$1/2$ &  \rule{0pt}{4mm}$({\bf 2_a})$ $({\bf 1_3})$  \\
$-1/2$ & $({\bf 2_a})$ $({\bf 1_3})$   \\
$3/2$ &  $({\bf 1_3})$  \\
$-3/2$ &   $({\bf 1_3})$\\
 & \\
\hline
\multicolumn{2}{|c|}{ \rule{0pt}{4mm}2-boundary chain, $b=1$ case}
\\
\hline
\multicolumn{2}{|c|}{\rule{0pt}{4mm}$({\bf 2_b\cdot 1_3\cdot 1_4\cdot
    1_5})$ $({\bf 1_4})^3$} \\
\hline
\multicolumn{2}{|c|}{\rule{0pt}{4mm}2-boundary chain, $b=0$ case} \\
\hline
\multicolumn{2}{|c|}{\rule{0pt}{4mm}$({\bf 2_c\cdot 2_d})^2$ } \\
\hline
\end{tabular}
\begin{tabular}{|r|l|}
\multicolumn{2}{c}{$\bf L=4$} \\
\hline
$S$ & \multicolumn{1}{|c|}{\rule{0pt}{4mm}open chain} \\
\hline
0 & \rule{0pt}{4mm}$({\bf 2_a})$ \\
1 & $({\bf 2_a})$ $({\bf 1_3})$ \\
2 & $({\bf 1_3})$\\
\hline
$m$ & \multicolumn{1}{|c|}{ \rule{0pt}{4mm}1-boundary chain} \\
\hline
 $0$ & \rule{0pt}{4mm}$({\bf 2_b\cdot 1_3\cdot 1_4\cdot 1_5})$$({\bf 1_4})$\\
$1$ &  $({\bf 2_c\cdot 2_d})$  \\
$-1$ & $({\bf 2_c\cdot 2_d})$   \\
$2$ &  $({\bf 1_4})$  \\
$-2$ &   $({\bf 1_4})$\\
\hline
\multicolumn{2}{|c|}{ \rule{0pt}{4mm}2-boundary chain, $b=1$ case} \\
\hline
\multicolumn{2}{|c|}{\rule{0pt}{4mm} --------} \\
\hline
\multicolumn{2}{|c|}{\rule{0pt}{4mm} 2-boundary chain, $b=0$ case} \\
\hline
\multicolumn{2}{|c|}{\rule{0pt}{4mm}
$({\bf 2_e\cdot 2_f\cdot 1_5})$ $({\bf 2_g\cdot 1_4\cdot 1_6})^2$
$({\bf 1_5})^3$ } \\
\hline
\end{tabular}

\noindent
\begin{tabular}{|r|l|}
\multicolumn{2}{c}{$\bf L=5$} \\
\hline
$S$ & \multicolumn{1}{|c|}{ \rule{0pt}{4mm} open chain} \\
\hline
\rule{0pt}{4mm}
1/2 & $({\bf 2_b\cdot 1_3\cdot 1_4\cdot 1_5})$ \\
3/2 & $({\bf 2_c\cdot 2_d})$ \\
5/2 & $({\bf 1_4})$\\ & \\
\hline
$m$ & \multicolumn{1}{|c|}{\rule{0pt}{4mm} 1-boundary chain} \\
\hline
\rule{0pt}{4mm}
$1/2$ & $({\bf 2_e\cdot 2_f\cdot 1_5})$ $({\bf 2_g\cdot 1_4\cdot 1_6})$
$({\bf 1_5})$ \\
$-1/2$ &  $({\bf 2_e\cdot 2_f\cdot 1_5})$ $({\bf 2_g\cdot 1_4\cdot 1_6})$
$({\bf 1_5})$ \\
$3/2$ &  $({\bf 2_g\cdot 1_4\cdot 1_6})$ $({\bf 1_5})$ \\
$-3/2$ &  $({\bf 2_g\cdot 1_4\cdot 1_6})$ $({\bf 1_5})$ \\
$5/2$&  $({\bf 1_5})$  \\
$-5/2$&  $({\bf 1_5})$  \\ & \\
\hline
\multicolumn{2}{|c|}{ \rule{0pt}{4mm} 2-boundary chain, $b=1$ case} \\
\hline
\multicolumn{2}{|c|}{\rule{0pt}{4mm}
$({\bf 3_a\cdot 3_b})^4$ $({\bf 3_c\cdot 3_d\cdot 1_5\cdot 1_7})$
} \\
\hline
\multicolumn{2}{|c|}{\rule{0pt}{4mm} 2-boundary chain, $b=0$ case} \\
\hline
\multicolumn{2}{|c|}{\rule{0pt}{4mm}
$({\bf 3_e\cdot 3_f\cdot 3_g\cdot 3_h\cdot 1_6^{\rm 2}})^2$
$({\bf 1_6})^4$
} \\ \multicolumn{2}{|c|}{} \\
\hline
\end{tabular}
\begin{tabular}{|r|l|}
\multicolumn{2}{c}{$\bf L=6$} \\
\hline
$S$ & \multicolumn{1}{|c|}{\rule{0pt}{4mm}open chain} \\
\hline
0 & \rule{0pt}{4mm}$({\bf 2_e\cdot 2_f\cdot 1_5})$  \\
1 & $({\bf 2_e\cdot 2_f\cdot 1_5})$ $({\bf 2_g\cdot 1_4\cdot 1_6})$ \\
2 & $({\bf 2_g\cdot 1_4\cdot 1_6})$ $({\bf 1_5})$\\
3 & $({\bf 1_5})$\\
\hline
$m$ & \multicolumn{1}{|c|}{ \rule{0pt}{4mm}1-boundary chain} \\
\hline
$0$ & \rule{0pt}{4mm}$({\bf 3_a\cdot 3_b})^2$
 $({\bf 3_c\cdot 3_d\cdot 1_5\cdot 1_7})$ \\
$1$ & $({\bf 3_e\cdot 3_f\cdot 3_g\cdot 3_h\cdot 1_6^{\rm 2}})$$({\bf 1_6})$ \\
$-1$ & $({\bf 3_e\cdot 3_f\cdot 3_g\cdot 3_h\cdot 1_6^{\rm 2}})$$({\bf 1_6})$ \\
$2$ &  $({\bf 3_a\cdot 3_b})$\\
$-2$ &  $({\bf 3_a\cdot 3_b})$ \\
$3$ & $({\bf 1_6})$\\
$-3$ & $({\bf 1_6})$\\
\hline
\multicolumn{2}{|c|}{ \rule{0pt}{4mm}2-boundary chain, $b=1$ case} \\
\hline
\multicolumn{2}{|c|}{\rule{0pt}{4mm}
------} \\
\hline
\multicolumn{2}{|c|}{\rule{0pt}{4mm} 2-boundary chain, $b=0$ case} \\
\hline
\multicolumn{2}{|c|}{\rule{0pt}{4mm}
$({\bf 4_a\cdot 2_h})^4$
 $({\bf 4_d\cdot 4_e\cdot 4_f\cdot 1_7^{\rm 2}})^2$}\\
\multicolumn{2}{|c|}{
$({\bf 4_b\cdot 4_c})$ $({\bf 1_7})^4$ } \\
\hline
\end{tabular}

\end{center}

Here\vspace{-2mm}
\be
\begin{array}{l}
\begin{array}{lll}
{\bf 1_n} := x-n, &  & \\[2mm]
{\bf 2_a} := x^2 - 6x + 7,\qquad\quad & {\bf 2_b} :=x^2 - 8x + 11,\qquad\quad &
{\bf 2_c} :=x^2 - 7x + 11, \\
{\bf 2_d} :=x^2 - 9x + 19,& {\bf 2_e} :=x^2 - 8x + 13, &
{\bf 2_f} :=x^2 - 12x + 33, \\
{\bf 2_g} :=x^2 - 10x + 22  , & {\bf 2_h} :=x^2 - 14x + 47  , &
\end{array}
\\[12mm]
\begin{array}{ll}
{\bf 3_a} :=x^3 - 17x^2 + 94x - 169 , &
{\bf 3_b} :=x^3 - 19x^2 + 118x - 239 ,\\
{\bf 3_c} :=x^3 - 17x^2 + 87x - 127 , &
{\bf 3_d} :=x^3 - 19x^2 + 111x - 197 ,\\
{\bf 3_e} :=x^3 - 16x^2 + 83x - 139 , \qquad\quad&
{\bf 3_f} :=x^3 - 18x^2 + 101x - 167 ,\\
{\bf 3_g} :=x^3 - 18x^2 + 101x - 181 , &
{\bf 3_h} :=x^3 - 20x^2 + 131x - 281 ,
\end{array}
\\[12mm]
\begin{array}{l}
{\bf 4_a} :=x^4 - 28x^3 + 290x^2 - 1316x + 2207, \\
{\bf 4_b}:=x^4 - 28x^3 + 282x^2 - 1188x + 1697, \\
{\bf 4_c} :=x^4 - 28x^3 + 282x^2 - 1220x + 1921, \\
{\bf 4_d}:=x^4 - 28x^3 + 286x^2 - 1252x + 1951, \\
{\bf 4_e} :=x^4 - 28x^3 + 286x^2 - 1260x + 2017, \\
{\bf 4_f}:=x^4 - 28x^3 + 286x^2 - 1268x + 2063,
\end{array}
\end{array}
\ee

\end{document}